%% file: paper.tex
\documentclass[linenumbers]{aastex631}

\begin{document}

\title{All-sky neutrino point-source search with IceCube combined track and cascade data}

\date{\today}
\affiliation{III. Physikalisches Institut, RWTH Aachen University, D-52056 Aachen, Germany}
\affiliation{Department of Physics, University of Adelaide, Adelaide, 5005, Australia}
\affiliation{Dept. of Physics and Astronomy, University of Alaska Anchorage, 3211 Providence Dr., Anchorage, AK 99508, USA}
\affiliation{School of Physics and Center for Relativistic Astrophysics, Georgia Institute of Technology, Atlanta, GA 30332, USA}
\affiliation{Dept. of Physics, Southern University, Baton Rouge, LA 70813, USA}
\affiliation{Dept. of Physics, University of California, Berkeley, CA 94720, USA}
\affiliation{Lawrence Berkeley National Laboratory, Berkeley, CA 94720, USA}
\affiliation{Institut f{\"u}r Physik, Humboldt-Universit{\"a}t zu Berlin, D-12489 Berlin, Germany}
\affiliation{Fakult{\"a}t f{\"u}r Physik {\&} Astronomie, Ruhr-Universit{\"a}t Bochum, D-44780 Bochum, Germany}
\affiliation{Universit{\'e} Libre de Bruxelles, Science Faculty CP230, B-1050 Brussels, Belgium}
\affiliation{Vrije Universiteit Brussel (VUB), Dienst ELEM, B-1050 Brussels, Belgium}
\affiliation{Dept. of Physics, Simon Fraser University, Burnaby, BC V5A 1S6, Canada}
\affiliation{Department of Physics and Laboratory for Particle Physics and Cosmology, Harvard University, Cambridge, MA 02138, USA}
\affiliation{Dept. of Physics, Massachusetts Institute of Technology, Cambridge, MA 02139, USA}
\affiliation{Dept. of Physics and The International Center for Hadron Astrophysics, Chiba University, Chiba 263-8522, Japan}
\affiliation{Department of Physics, Loyola University Chicago, Chicago, IL 60660, USA}
\affiliation{Dept. of Physics and Astronomy, University of Canterbury, Private Bag 4800, Christchurch, New Zealand}
\affiliation{Dept. of Physics, University of Maryland, College Park, MD 20742, USA}
\affiliation{Dept. of Astronomy, Ohio State University, Columbus, OH 43210, USA}
\affiliation{Dept. of Physics and Center for Cosmology and Astro-Particle Physics, Ohio State University, Columbus, OH 43210, USA}
\affiliation{Niels Bohr Institute, University of Copenhagen, DK-2100 Copenhagen, Denmark}
\affiliation{Dept. of Physics, TU Dortmund University, D-44221 Dortmund, Germany}
\affiliation{Dept. of Physics and Astronomy, Michigan State University, East Lansing, MI 48824, USA}
\affiliation{Dept. of Physics, University of Alberta, Edmonton, Alberta, T6G 2E1, Canada}
\affiliation{Erlangen Centre for Astroparticle Physics, Friedrich-Alexander-Universit{\"a}t Erlangen-N{\"u}rnberg, D-91058 Erlangen, Germany}
\affiliation{Physik-department, Technische Universit{\"a}t M{\"u}nchen, D-85748 Garching, Germany}
\affiliation{D{\'e}partement de physique nucl{\'e}aire et corpusculaire, Universit{\'e} de Gen{\`e}ve, CH-1211 Gen{\`e}ve, Switzerland}
\affiliation{Dept. of Physics and Astronomy, University of Gent, B-9000 Gent, Belgium}
\affiliation{Dept. of Physics and Astronomy, University of California, Irvine, CA 92697, USA}
\affiliation{Karlsruhe Institute of Technology, Institute for Astroparticle Physics, D-76021 Karlsruhe, Germany}
\affiliation{Karlsruhe Institute of Technology, Institute of Experimental Particle Physics, D-76021 Karlsruhe, Germany}
\affiliation{Dept. of Physics, Engineering Physics, and Astronomy, Queen's University, Kingston, ON K7L 3N6, Canada}
\affiliation{Department of Physics {\&} Astronomy, University of Nevada, Las Vegas, NV 89154, USA}
\affiliation{Nevada Center for Astrophysics, University of Nevada, Las Vegas, NV 89154, USA}
\affiliation{Dept. of Physics and Astronomy, University of Kansas, Lawrence, KS 66045, USA}
\affiliation{Centre for Cosmology, Particle Physics and Phenomenology - CP3, Universit{\'e} catholique de Louvain, Louvain-la-Neuve, Belgium}
\affiliation{Department of Physics, Mercer University, Macon, GA 31207-0001, USA}
\affiliation{Dept. of Astronomy, University of Wisconsin{\textemdash}Madison, Madison, WI 53706, USA}
\affiliation{Dept. of Physics and Wisconsin IceCube Particle Astrophysics Center, University of Wisconsin{\textemdash}Madison, Madison, WI 53706, USA}
\affiliation{Institute of Physics, University of Mainz, Staudinger Weg 7, D-55099 Mainz, Germany}
\affiliation{Department of Physics, Marquette University, Milwaukee, WI 53201, USA}
\affiliation{Institut f{\"u}r Kernphysik, Universit{\"a}t M{\"u}nster, D-48149 M{\"u}nster, Germany}
\affiliation{Bartol Research Institute and Dept. of Physics and Astronomy, University of Delaware, Newark, DE 19716, USA}
\affiliation{Dept. of Physics, Yale University, New Haven, CT 06520, USA}
\affiliation{Columbia Astrophysics and Nevis Laboratories, Columbia University, New York, NY 10027, USA}
\affiliation{Dept. of Physics, University of Oxford, Parks Road, Oxford OX1 3PU, United Kingdom}
\affiliation{Dipartimento di Fisica e Astronomia Galileo Galilei, Universit{\`a} Degli Studi di Padova, I-35122 Padova PD, Italy}
\affiliation{Dept. of Physics, Drexel University, 3141 Chestnut Street, Philadelphia, PA 19104, USA}
\affiliation{Physics Department, South Dakota School of Mines and Technology, Rapid City, SD 57701, USA}
\affiliation{Dept. of Physics, University of Wisconsin, River Falls, WI 54022, USA}
\affiliation{Dept. of Physics and Astronomy, University of Rochester, Rochester, NY 14627, USA}
\affiliation{Department of Physics and Astronomy, University of Utah, Salt Lake City, UT 84112, USA}
\affiliation{Dept. of Physics, Chung-Ang University, Seoul 06974, Republic of Korea}
\affiliation{Oskar Klein Centre and Dept. of Physics, Stockholm University, SE-10691 Stockholm, Sweden}
\affiliation{Dept. of Physics and Astronomy, Stony Brook University, Stony Brook, NY 11794-3800, USA}
\affiliation{Dept. of Physics, Sungkyunkwan University, Suwon 16419, Republic of Korea}
\affiliation{Institute of Physics, Academia Sinica, Taipei, 11529, Taiwan}
\affiliation{Dept. of Physics and Astronomy, University of Alabama, Tuscaloosa, AL 35487, USA}
\affiliation{Dept. of Astronomy and Astrophysics, Pennsylvania State University, University Park, PA 16802, USA}
\affiliation{Dept. of Physics, Pennsylvania State University, University Park, PA 16802, USA}
\affiliation{Dept. of Physics and Astronomy, Uppsala University, Box 516, SE-75120 Uppsala, Sweden}
\affiliation{Dept. of Physics, University of Wuppertal, D-42119 Wuppertal, Germany}
\affiliation{Deutsches Elektronen-Synchrotron DESY, Platanenallee 6, D-15738 Zeuthen, Germany}

\author[0000-0001-6141-4205]{R. Abbasi}
\affiliation{Department of Physics, Loyola University Chicago, Chicago, IL 60660, USA}

\author[0000-0001-8952-588X]{M. Ackermann}
\affiliation{Deutsches Elektronen-Synchrotron DESY, Platanenallee 6, D-15738 Zeuthen, Germany}

\author{J. Adams}
\affiliation{Dept. of Physics and Astronomy, University of Canterbury, Private Bag 4800, Christchurch, New Zealand}

\author[0000-0002-9714-8866]{S. K. Agarwalla}
\altaffiliation{also at Institute of Physics, Sachivalaya Marg, Sainik School Post, Bhubaneswar 751005, India}
\affiliation{Dept. of Physics and Wisconsin IceCube Particle Astrophysics Center, University of Wisconsin{\textemdash}Madison, Madison, WI 53706, USA}

\author[0000-0003-2252-9514]{J. A. Aguilar}
\affiliation{Universit{\'e} Libre de Bruxelles, Science Faculty CP230, B-1050 Brussels, Belgium}

\author[0000-0003-0709-5631]{M. Ahlers}
\affiliation{Niels Bohr Institute, University of Copenhagen, DK-2100 Copenhagen, Denmark}

\author[0000-0002-9534-9189]{J.M. Alameddine}
\affiliation{Dept. of Physics, TU Dortmund University, D-44221 Dortmund, Germany}

\author[0009-0001-2444-4162]{S. Ali}
\affiliation{Dept. of Physics and Astronomy, University of Kansas, Lawrence, KS 66045, USA}

\author{N. M. Amin}
\affiliation{Bartol Research Institute and Dept. of Physics and Astronomy, University of Delaware, Newark, DE 19716, USA}

\author[0000-0001-9394-0007]{K. Andeen}
\affiliation{Department of Physics, Marquette University, Milwaukee, WI 53201, USA}

% ...

\collaboration{427}{IceCube Collaboration}



\begin{abstract}

Despite extensive efforts, discovery of high-energy astrophysical neutrino sources remains elusive. We present an event-level simultaneous maximum likelihood analysis of tracks and cascades using IceCube data collected from 04/06/2008 to 05/23/2022 to search the whole sky for neutrino sources and, using a source catalog, for coincidence of neutrino emission with gamma-ray emission. This is the first time a simultaneous fit of different detection channels is used to conduct a time-integrated all-sky scan with IceCube. Combining all-sky tracks, with superior pointing-power and sensitivity in the northern sky, with all-sky cascades, with good energy-resolution and sensitivity in the southern sky, we have developed the most sensitive point-source search to date by IceCube which targets the entire sky. The most significant point in the northern sky aligns with NGC 1068, a Seyfert II galaxy, which, from the catalog search, shows a 3.5$\sigma$ excess over background after accounting for trials. The most significant point in the southern sky does not align with any source in the catalog and is not significant after accounting for trials. A search for the single most significant Gaussian flare at the locations of NGC 1068, PKS 1424+240, and the southern highest significance point shows results consistent with expectations for steady emission. Notably, this is the first time that a flare shorter than four years has been excluded as being responsible for NGC 1068’s emergence as a neutrino source. Our results show that combining tracks and cascades when conducting neutrino source searches improves sensitivity and can lead to new discoveries.

\end{abstract}

\keywords{Neutrino astronomy(1100) --- Neutrino telescopes(1105) --- High energy astrophysics(739) --- Particle astrophysics(96)}


\section{Introduction} \label{sec:intro}

Nearly massless and electrically neutral fundamental particles called neutrinos permeate our universe in three different flavors: the electron neutrino ($\nu_e$), the muon neutrino ($\nu_\mu$), and the tau neutrino ($\nu_\tau$). Although $\nu_e$ were first discovered in 1956 as by-products of radioactive decay in nuclear reactors \citep{cowan_1956}, it was later predicted that neutrinos could also be produced through cosmic-ray interactions near astrophysical sources \citep{Spiering_2012_review}. This suggested that neutrinos could potentially be used as astrophysical messengers, helping us locate high-energy astrophysical events such as active galactic nuclei and gamma-ray bursts. Unlike photons, which interact with the cosmic microwave background (CMB) and extragalactic background light at high energies \citep{Biteau_2022_Cosmology}, and cosmic rays (high-energy charged nuclei), which can be deflected by magnetic fields and attenuated at high energies ($10^{19}$ eV) by the CMB \citep{Watson_2014}, high-energy neutrinos can travel cosmological distances in a straight line without interacting or being deflected. This allows us to investigate the most distant and energetic astrophysical events. Although neutrinos are very difficult to detect due to their small cross sections, their interaction probability increases with energy \citep{Formaggio_2012}. This phenomenon is an important reason why the IceCube detector is most sensitive to neutrinos in the energy range of TeV to PeV, where detection probability is highest. 

\subsection{IceCube Neutrino Observatory}

Located at the Amundsen-Scott South Pole Station, the IceCube Neutrino Observatory was built to detect the Cherenkov radiation that follows after neutrinos interact with the water molecules that make up the Antarctic ice \citep{detector_paper}. IceCube  consists of a cubic-kilometer array of 5160 digital optical modules (DOMs) spread over 86 vertical cables (strings). Each DOM contains a 10" photomultiplier tube (PMT) which can detect the arrival time and intensity of photons from Cherenkov radiation \citep{pmt_paper} as well as an onboard data acquisition system \citep{daq_paper}. The incoming direction and energy of neutrino events can then be reconstructed using this information. 

The pattern of collected light varies based on the neutrino flavor and the type of interaction—charged-current (CC) or neutral-current (NC). The two main categories of event patterns are tracks and cascades. Tracks, produced predominantly by the outgoing muon in $\nu_\mu$ CC interactions, are linear in shape and usually extend for many kilometers, allowing detection even if the interaction vertex is outside the detector. Cascades appear nearly spherical in shape and are produced predominantly by CC interactions of $\nu_e$ and $\nu_\tau$ but also by NC interactions from all neutrino flavors \citep{Cascade2017, DNNCascades_2023}. 

\section{Dataset Construction}
\subsection{Track Events}
Prior searches by IceCube for neutrino point-sources have relied primarily on track events.  Historically, tracks have been favored over cascades due to their small angular errors (less than 1$^\circ$ at greater than TeV energies \citep{10yrPSTracks}) and their high detection rate of about 2 mHz, allowing for a very robust sample. In 2017, a high-energy neutrino track event was detected by IceCube and traced back to the blazar TXS 0506+056 \citep{realtime_alert, TXS_alert, TXS_Flare}, providing the first evidence of neutrino emission from a non-stellar astrophysical source. In 2022, track events were used as evidence for neutrino emission from the active galaxy NGC 1068 \citep{NorthernTracks}. 

Although tracks have small angular uncertainties, they are contaminated with background from atmospheric $\mu$ and $\nu$, with 100 million $\mu$ detected for every astrophysical $\nu$ at trigger level. In the northern sky, the Earth acts as a filter, absorbing and reducing the amount of $\mu$ background. Since there is less material between the atmosphere and the IceCube detector in the southern sky, the rate of atmospheric $\mu$ detected is orders of magnitude higher than atmospheric $\nu$ \citep{Instrumentation}. To formalize this division, IceCube defines its operational horizon at a declination of $\delta = -5^\circ$, corresponding to an overburden of at least 12 km of ice. This boundary, slightly south of the geometric horizon at $\delta = 0^\circ$, serves as the working separation between northern and southern skies for point-source searches. Even though the Earth filters out most of the atmospheric $\mu$ from the northern sky, southern sky atmospheric $\mu$ can be poorly reconstructed, also creating $\mu$ background in the northern sky \citep{10yrPSTracks}. Atmospheric $\nu$ contamination when CC $\nu_\mu$ interactions produce $\mu$ occurs in both the northern and southern skies.

To mitigate the large atmospheric muon background in the southern sky, low-energy track events from the southern sky (with reconstructed energies less than 10 TeV) are cut \citep{PSTracksCuts, IC40_2011}. These cuts are based on a single unbroken power law of $E^{-\gamma}$, where $\gamma$ is the spectral index. This power-law energy spectrum describes how the number of neutrinos per unit energy from a source decreases as energy increases, with $\gamma$ defining the rate of decrease. The energy of atmospheric neutrinos follows a ``softer" power-law of $E^{-3.7}$, while the energy of galactic and extragalactic neutrinos follow a much ``harder" energy spectrum \citep{Aartsen_2019}. While atmospheric neutrino backgrounds exist in both the northern and southern sky, the additional overwhelming atmospheric muon contamination in the southern sky necessitates these aggressive energy cuts. Consequently, IceCube track data is more sensitive to point-like sources in the northern sky than in the southern sky. The track dataset used in this analysis adds approximately four additional years to the previous all-sky point-source search conducted with the point-source tracks data sample (PSTracks) \citep{10yrPSTracks}, totaling about 1.5 million events in about 14 years of exposure.

\subsection{Cascade Events}
Recently, cascades have been used to find evidence of neutrino emission from the galactic plane \citep{DNNCascades_2023}, showing that cascades can also be used to find sources. Cascades have larger directional uncertainties, on the order of 10$^\circ$ at energies above 10 TeV \citep{Cascade2017}, compared to tracks. However, because cascades are mostly contained within the detector, a more accurate neutrino energy measurement can be made \citep{EnergyReco}.
Since cascade detection requires interaction vertices to be in or near the detector, the cascade dataset used in this analysis is smaller than the track dataset, containing only about 60,000 events over 10 years. The cascade dataset used in this analysis is the same as that used in \cite{DNNCascades_2023}, which was used for the first observation of neutrino emission from the Galactic plane. These events, however, have higher signal purity in the southern sky compared to tracks. The background for cascades comes from NC interactions of all-flavor atmospheric neutrinos as well as CC interactions of $\nu_\tau$ and $\nu_e$. Since cascades are not produced by atmospheric muons or $\nu_\mu$ CC interactions, fewer energy cuts are necessary to remove background. The selection of cascades thus has an order of magnitude fewer background neutrinos at TeV energies compared to tracks. This reduced background also lowers the median energy of cascades to about 1 TeV across the entire sky. Consequently, cascades have better sensitivity to point-like sources in the southern sky compared to tracks.
\subsection{Combining Tracks and Cascades}
Combining the 14 years of track data with the 10 years of cascade data will result in IceCube’s best sensitivity to point-like sources across the entire sky. With tracks leading the point-source sensitivity in the northern sky and cascades leading the point-source sensitivity in the southern sky, this combined dataset will fully exploit the benefits of both morphologies. Although new for the IceCube Collaboration, the idea of combining the track and cascade data channels is not new and has been done by the ANTARES Collaboration \citep{Antares}.

The events in both the track and cascade component datasets are assigned unique identifiers called event IDs. Due to some events having overlapping classifications after the event selection process, 703 events in each component dataset were found to have matching event IDs. These events were manually removed from the tracks dataset but retained in the cascades dataset. These overlapping events account for only 0.05\% of the tracks dataset and were kept in the smaller cascades dataset to maintain robustness.

An investigation into Monte Carlo (MC) simulation overlap between component datasets was also conducted. This test was done by applying the tracks event selection pipeline to cascades MC and vice versa. Any MC events that survived through the selection process were manually removed and sensitivities were re-calculated using these new MC events. This test resulted in sensitivities which were maximally different by $\sim3\%$ and minimally different by $\sim0.5\%$. Due to the minimal changes in sensitivity, overlapping MC events were not removed from the component datasets for this analysis. 

\begin{figure}[h!]
\centering
\includegraphics[width=.8\linewidth]{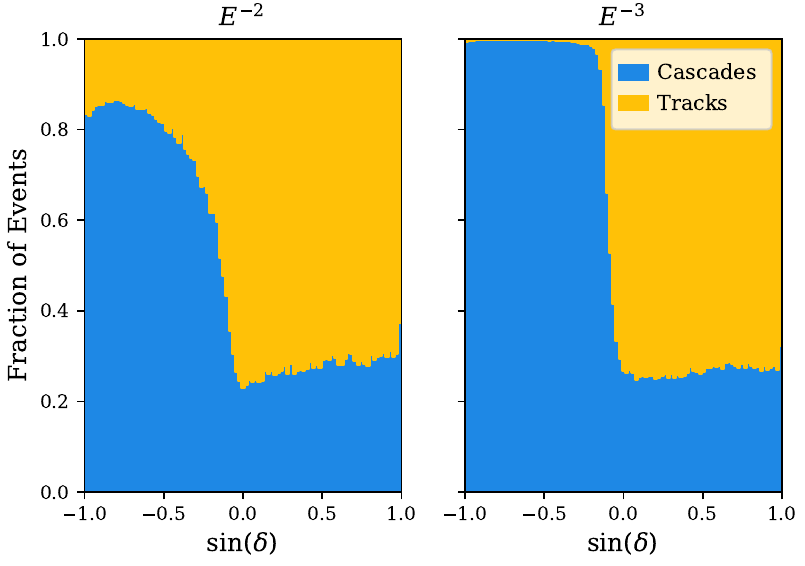}
\caption{\textbf{Fraction of expected signal events by event type and declination.} The relative fraction of events expected from cascades and tracks as a function of declination assuming an $E^{-2}$ spectrum (left) and $E^{-3}$ spectrum (right).
\label{fig:rel_sig_acc}}
\end{figure}

Table \ref{tab:datasamples} shows the livetime in days, number of events, and the start/end dates for all samples in the combined tracks and cascades dataset. Fig.~\ref{fig:rel_sig_acc} shows the relative fraction of signal events expected from cascades and tracks over the entire sky. For both an $E^{-2}$ and $E^{-3}$ spectrum, cascade events dominate in the southern sky, while track events dominate in the northern sky. Our sensitivity to point-sources improves the most at the declination where signal events are expected to come from tracks and cascades equally (50\% from each). This critical declination depends on the assumed energy spectrum. For an $E^{-3}$ spectrum, the critical declination occurs at approximately $-5^\circ$. This declination is the definitional boundary between `upgoing' and `downgoing' events. For an $E^{-2}$ spectrum, the critical declination shifts slightly south to about $-8^\circ$. This difference in critical declination between the two assumed spectra arises because the softer $E^{-3}$ spectrum shifts events to lower energies, where the cascades dataset's lower energy threshold enables detection of more events from the southern sky compared to the tracks dataset. 

\begin{figure}[h!]
\includegraphics[width=\linewidth]{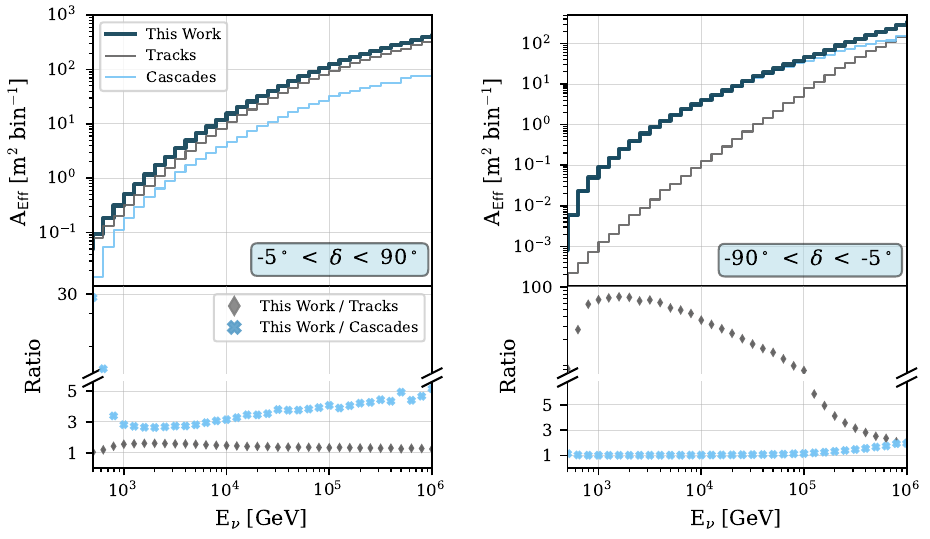}
\caption{\textbf{Effective Area Comparison.} The effective area in $m^2$ (averaged over solid angle) of all-flavor neutrinos per energy bin for tracks (this work), cascades \citep{DNNCascades_2023}, and the combined sample (this work). Effective areas are averaged over solid angle in the declination range of $-5^\circ$ to 90$^\circ$  for the northern sky (left) and in the range of $-90^\circ$ to $-5^\circ$ for the southern sky (right). Each decade in energy contains 10 bins. The ratio of the combined sample effective area to each component effective area is shown for each sky. 
\label{fig:eff_area}}
\end{figure}

\begin{figure}[h!]
\includegraphics[width=\linewidth]{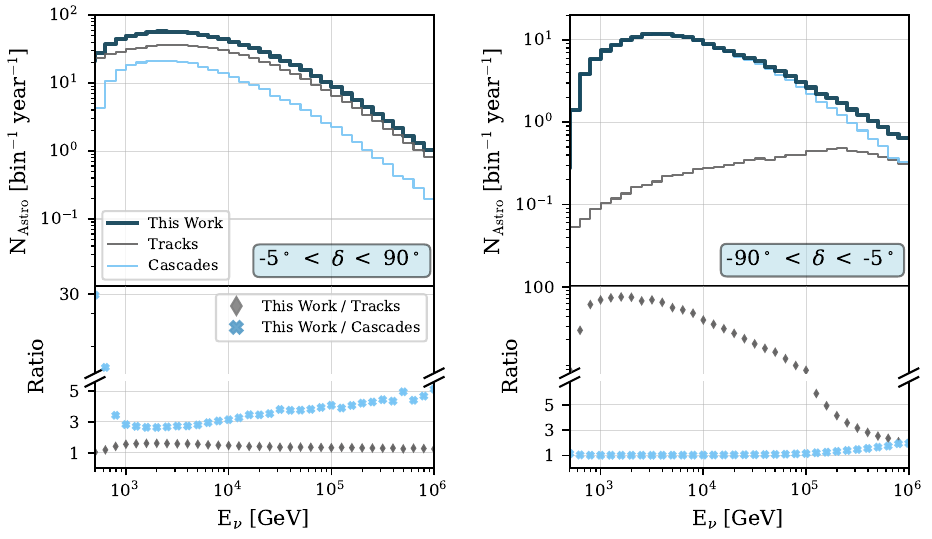}
\caption{\textbf{Expected Number of Astrophysical Neutrinos Comparison.} The number of expected all-flavor astrophysical neutrinos (N$_{\mathrm{Astro}}$) per energy bin per year with respect to energy for tracks (this work), cascades \citep{DNNCascades_2023}, and the combined sample (this work) assuming an astrophysical spectrum of $E^{-2.58}$ \citep{ESTES_astro_flux} for the northern sky (left) and the southern sky (right). Each decade in energy contains 10 bins. The ratio of the combined sample N$_{\mathrm{Astro}}$ to each component N$_{\mathrm{Astro}}$ is shown for each sky.}

\label{fig:n_astro}
\end{figure}

Figs.~\ref{fig:eff_area} and~\ref{fig:n_astro} show the effective area and expected number of astrophysical neutrino events respectively in the northern and southern sky for tracks only, cascades only, and combined tracks and cascades. In the northern sky, both the effective area and the expected number of astrophysical neutrino events of the combined tracks and cascades sample is higher than either tracks or cascades on their own in the range of 1 TeV to 1 PeV and is about twice as high as either between 1 TeV and 10 TeV. In the southern sky, both the combined effective area and the expected number of astrophysical neutrino events are slightly higher than that of cascades until energies higher than 100 TeV where the combined effective area and the number of events shifts to being about twice as high as either individual dataset. The improvement in effective area and expected number of events due to combining datasets is larger in the northern sky than in the southern sky. 

\begin{deluxetable}{lccccc}
\tabletypesize{\small}
\tablewidth{0pt}
\tablecaption{Data Samples \label{tab:datasamples}}
\tablehead{
\colhead{String Config./Data Sample} & \colhead{Livetime (Days)} & \colhead{Number of Events} & \colhead{Start Date} & \colhead{End Date} & \colhead{Ref.}
}
\startdata
IC40 Tracks & 376.4 & 36900 & 04/06/2008 & 05/20/2009 & \cite{IC40_2011}\\
\hline
IC59 Tracks & 353.6 & 107011 & 05/20/2009 & 05/31/2010 & \cite{3yrPSTracks}\\
\hline
IC79 Tracks & 312.8 & 101956 & 06/01/2010 & 05/13/2011 & \cite{IC79_2011}\\
\hline
IC86 Tracks & 3920.6 & 1397502 & 05/13/2011 & 05/23/2022 & \cite{10yrPSTracks}\\
\hline
IC86 Cascades & 3519.1 & 59592 & 05/13/2011 & 05/27/2021 & \cite{DNNCascades_2023}\\
\enddata
\tablecomments{\textbf{Dataset Properties.} Data sample (string configuration and data channel), livetime, number of events, start and end date, and published reference for each sample used in this work. }
\end{deluxetable}

\section{Time-Integrated All-Sky Point-Source Search}
\subsection{Analysis Method}

This analysis utilizes the above-mentioned newly combined dataset to identify the most significant astrophysical neutrino point-sources across the entire sky. This is a time-integrated analysis (covering the entire livetime of the detector) that employs the maximum likelihood technique described by \cite{3yrPSTracks}. This method provides a way to combine datasets regardless of varying event rates, signal purity, MC simulations, or expected flavors. The likelihood function, $\mathcal{L}$, is defined in Eq.~\ref{base_likelihood}. This likelihood is maximized by fitting for $n_s$ (mean number of signal events) and $\gamma$, and is composed of signal and background probability density functions (PDFs). Each PDF is composed of a spatial component and an energy component, enabling the calculation of the probability of an event being either signal or background. The following equation shows the general form of the likelihood: 

\begin{equation} \label{base_likelihood}
\mathcal{L}(n_s,\gamma ) = \prod_{j}^{M} \prod_{i \in j}^{N}\frac{n_s^j}{N^j}\mathcal{S}(\textbf{x}_S, \textbf{x}_i^j, \sigma_i^j, E_i^j; \gamma) + \left(1 - \frac{n_s^j}{N^j}\right)\mathcal{B}(\sin \delta_i^j, E_i^j).
\end{equation}
As the best-fit $n_s$ increases, the first term of the likelihood (the signal weighting term) increases and the second term of the likelihood (the background weighting term) decreases. The fraction of signal events from each dataset, indexed by $j$, depends on $\gamma$ and declination, $\delta$, and is used to calculate the number of signal events per dataset, $n_s^j$. Examples of these fractions are shown in Fig.~\ref{fig:rel_sig_acc}. Thus, $\mathcal{L}$ is only a function of global $n_s$ and $\gamma$. The total number of events in each data sample is $N^j$, the signal PDF is $\mathcal{S}$, the background PDF is $\mathcal{B}$, the declination of the $i^\mathrm{th}$ event in the $j^{\mathrm{th}}$ sample is $\delta_i^j$, the source position of the event in right ascension ($\alpha$) and $\delta$ is $\textbf{x}_S$, the reconstructed direction of the neutrino event is $\textbf{x}_i^j$, and the angular uncertainty is $\sigma$ \citep{Braun_2008, TXS_Flare}. $\mathcal{S}$ and $\mathcal{B}$ are evaluated at the reconstructed position of the event. The spatial component of $\mathcal{S}$ is modeled using either a Gaussian or a von Mises-Fisher distribution, depending on the evaluated event's $\sigma$. For events with $\sigma \le 7^\circ$, the Gaussian form is appropriate, as it remains valid under the small-angle approximation. This form of $\mathcal{S}$ is shown below:
\begin{equation} \label{S_gaussian}
\mathcal{S} = \frac{1}{2\pi \sigma_i^2}e^\frac{-|\textbf{x}_S-\textbf{x}_i|^2}{2\sigma_i^2} \times \mathcal{E_S}(E_i, \sin\delta_i;\gamma).
\end{equation}
On the other hand, if $\sigma > 7^\circ$, the Gaussian approximation breaks down due to spherical geometry effects, and the von Mises-Fisher distribution is used instead. This form of $\mathcal{S}$ is shown below:
\begin{equation} \label{S_vMF}
\mathcal{S} = \frac{1}{4\pi \sigma_i^2\sinh{(\frac{1}{\sigma_i^2})}}e^\frac{\cos(|\textbf{x}_S-\textbf{x}_i|)}{\sigma_i^2} \times \mathcal{E_S}(E_i, \sin\delta_i;\gamma).
\end{equation}
The energy component ($\mathcal{E_S}$) of $\mathcal{S}$ accounts for the probability of seeing an event with energy $E_i$ at a declination of $\delta_i$ given a flux model with spectral index $\gamma$. The form of $\mathcal{B}$ is shown below: 
\begin{equation} \label{B}
\mathcal{B} = \mathcal{P_B}(\sin\delta_i) \times \mathcal{E_B}(E_i,\sin\delta_i).
\end{equation}
The spatial component ($\mathcal{P_B}$) of $\mathcal{B}$ is given by the event density per solid angle, normalized by the total number of events in the sample. The energy component ($\mathcal{E_B}$) of the $\mathcal{B}$ is taken from data and is measured by the fraction of events with energy $E_i$ at a declination of $\sin\delta_i$. 

The test statistic (TS) used for this analysis is shown in Eq.~\ref{test_stat} where $\hat{n}_s$ is the best-fit number of signal events and $\hat{\gamma}$ is the best-fit spectral index. As $\hat{n}_s$ becomes larger, $\mathcal{L}(\hat{n}_s,\hat{\gamma})$ becomes larger than the null hypothesis/background hypothesis $\mathcal{L}(n_s=0)$ and thus the TS increases. 
\begin{equation} \label{test_stat}
\text{TS} = -2 \log \left[ \frac{\mathcal{L}(n_s=0)}{\mathcal{L}(\hat{n}_s, \hat{\gamma})} \right].
\end{equation}

To simulate background, for use in TS calculations, background trials are generated by randomly scrambling the $\alpha$ of events in the actual data. The rotation of the Earth ensures that atmospheric backgrounds, isotropic astrophysical neutrino background, and other Earth-based sources of systematic uncertainty are uniformly distributed in $\alpha$. This makes it possible to generate realistic background trials from data while taking into account the systematic effects of the detector. This method generally eliminates any clustering signals, however, to accurately create these background trials, residual signal must be subtracted. This adjustment is necessary because, with the galactic plane covering a significant portion of some declination bands, randomizing in $\alpha$ no longer ensures that the events are background only but rather a mix of signal and background. To achieve this, previous IceCube all-sky point-source searches with cascades added a signal subtraction likelihood to Eq.~\ref{base_likelihood} \citep{7yrPSTracks,DNNCascades_2023}. To maintain consistency in the likelihoods throughout this analysis, signal subtraction is applied to both the cascades and tracks datasets. The signal subtraction likelihood is the following:

\begin{equation} \label{sig_sub}
\tilde{\mathcal{D}} =  \frac{n_s^j}{N^j}\tilde{\mathcal{S}}(\sin(\delta_i^j), E_i^j) + \left(1- \frac{n_s^j}{N^j}\right)\mathcal{B}(\sin(\delta_i^j), E_i^j).
\end{equation}
Solving for $\left(1-n_s^j/N^j\right)\mathcal{B}(\sin(\delta_i^j),E_i^j)$ in Eq.~\ref{sig_sub} and replacing it in Eq.~\ref{base_likelihood} gives us the equation below:
\begin{equation} \label{new_L}
\mathcal{L}(n_s, \gamma) = \prod_{j}^M \prod_{i \in j}^N \frac{n_s^j}{N^j}\mathcal{S}(\delta_i^j, \gamma, \sigma_i^j) + \tilde{\mathcal{D}} (\sin(\delta_i^j), E_i^j) - \frac{n_s^j}{N^j}\tilde{\mathcal{S}} (\sin(\delta_i^j), E_i^j).
\end{equation}

Here, $\tilde{\mathcal{D}}$ is the background expectation from scrambling and $\tilde{\mathcal{S}}$ is the $\alpha$-averaged expected number of signal events based on MC event acceptance to the source hypothesis which, in this case, is a point-source hypothesis.

Previous cascades-based point-source searches observed noise in the sensitivity distributions when background trials were constructed solely by randomizing in $\alpha$. This effect arises from the limited statistics of the cascades sample which reduce the number of independent scrambles possible. To mitigate this, an additional Gaussian randomization in $\delta$ with a Gaussian width of 3$^\circ$~\citep{DNNCascades_2023} is applied when generating background from the cascade sample. This procedure is only necessary and valid for cascades since the larger sample size of the tracks sample provides sufficient statistics with $\alpha$-randomization alone, and the large angular uncertainties of cascades validates the chosen declination scrambling width of 3$^\circ$. Consequently, this randomization in $\delta$ is maintained for background constructed from the cascades sample but is not used for background constructed from the tracks sample.

Declination-based signal is simulated to determine our analysis sensitivity and discovery potential across the sky. When combining tracks and cascades, we inject signal-like events (signal trials) from each dataset based on their fraction of expected events for a specified energy spectrum (see Fig. \ref{fig:rel_sig_acc}) around the declination of interest. Given tracks' finer angular resolution compared to cascades, MC events sampled from $\pm$3° in declination for tracks injections and $\pm$5° for cascades injections. Descriptions of the four tracks samples and the cascades sample used in the combined tracks and cascades dataset are shown in Table~\ref{tab:datasamples}.

\subsection{Sensitivity and Discovery Potential}

In order to characterize this analysis and allow for comparison to other analyses, sensitivities were calculated for $\gamma = 2$ and $\gamma = 3$ across the entire sky. Sensitivities are determined constructing signal trials at each declination and then identifying the number of injected signal events ($n_{\mathrm{inj}}$) and the corresponding flux required for 90\% of signal trials to yield a test statistic (TS) greater than the median TS derived from background trials. Additionally, the 5$\sigma$ discovery potential is calculated, defined as the $n_{\mathrm{inj}}$ and corresponding flux required for 50\% of signal trials to achieve a TS greater than the background TS distribution's 5$\sigma$ threshold. 

\begin{figure}[h!]
\includegraphics[width=\linewidth]{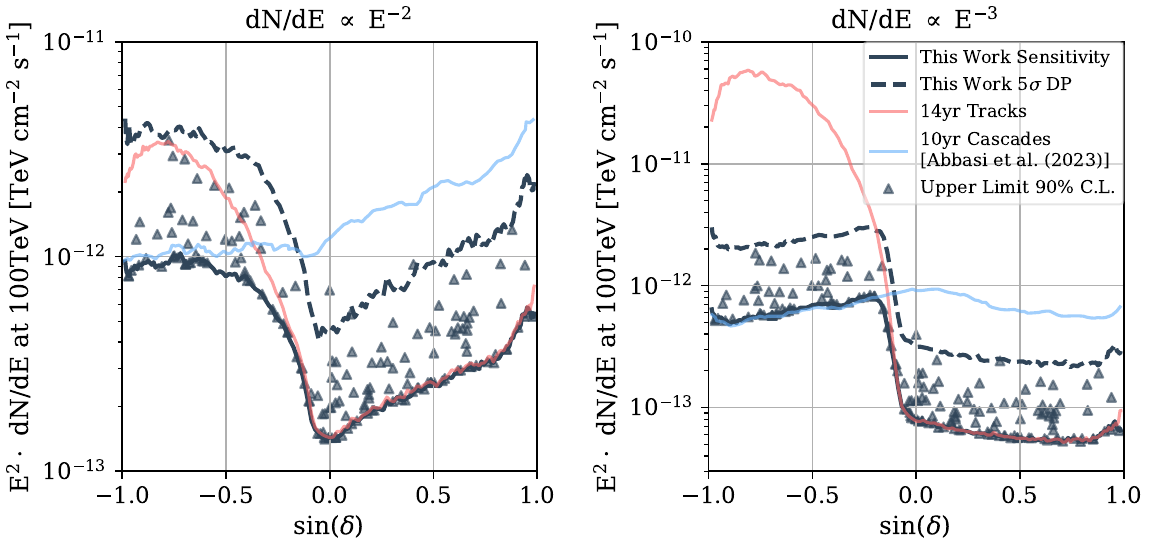}
\caption{\textbf{Source List Sensitivity, 5$\sigma$ Discovery Potential and Upper Limits.} 90\% C.L. median sensitivity to sources emitting an $E^{-2}$ spectrum (left) and $E^{-3}$ spectrum (right) as a function of source declination for cascades \citep{DNNCascades_2023} and tracks individually and for combined tracks and cascades. The 90\% C.L. upper-limits for the source catalog sources are shown assuming an $E^{-2}$ spectrum (left) and $E^{-3}$ spectrum (right). For sources with $\hat{n}_s=0$, the 90\% C.L. median sensitivity is used instead of the upper limit. The 5$\sigma$ discovery potential (DP) for combined tracks and cascades is also shown. $\frac{dN}{dE}$ is the per-flavor number of neutrinos ($N$) per neutrino energy ($E$) per area per time. 
\label{fig:all_sky_sens}}
\end{figure}

The sensitivities for the combined dataset and both components and the combined dataset's 5$\sigma$ discovery potential are shown in Fig.~\ref{fig:all_sky_sens} for $\gamma= 2$  and $\gamma= 3$. The 5$\sigma$ discovery potentials for tracks and cascades individually are shown in Fig.~\ref{fig:all_data_sens}. For $\gamma = 2$, the sensitivity is improved across the southern sky after combining the cascades and tracks datasets. The 5$\sigma$ discovery potential is similarly improved across the southern sky. For $\gamma= 3$, there is improvement by combining the datasets in both sensitivity and 5$\sigma$ discovery potential just below the horizon in the range of about sin($\delta$) = [$-0.19$, $-0.11$].

\subsection{Results}
\subsubsection{All-Sky Search}
\begin{figure}[h!]
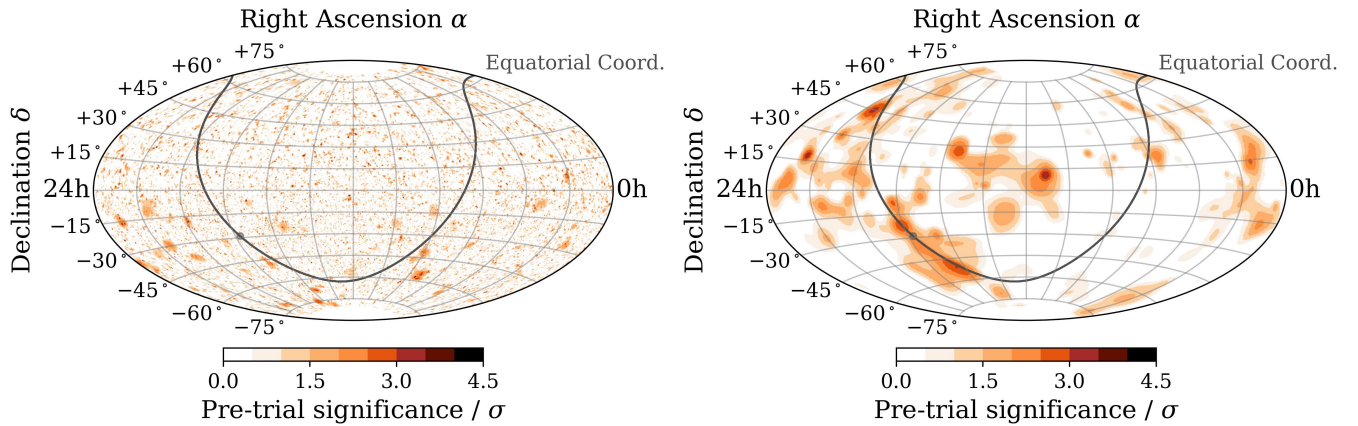

\includegraphics[width=.49\textwidth]{tracks_skymap.pdf}
\includegraphics[width=.49\textwidth]{cascades_skymap.pdf}
\caption{\textbf{Tracks and Cascades All-Sky Maps.} Best-fit pre-trial significance all-sky Aitoff projection maps using the 14 years of tracks from this work (left) and for comparison, previous 10 year cascade results (right) \citep{DNNCascades_2023} as a function of direction in equatorial coordinates (J2000 equinox). This all-sky tracks sky-map adds 3.6 years of data to the previous all-sky neutrino point-source search with tracks \citep{10yrPSTracks}. The solid gray line denotes the galactic plane with the dot representing the galactic center. The hottest northern and southern spots on both maps are in different locations as described in Table \ref{tab:all_sky_results}. 
\label{fig:sep_all_sky_scan}}
\end{figure}

\begin{figure}[h!]
\plotone{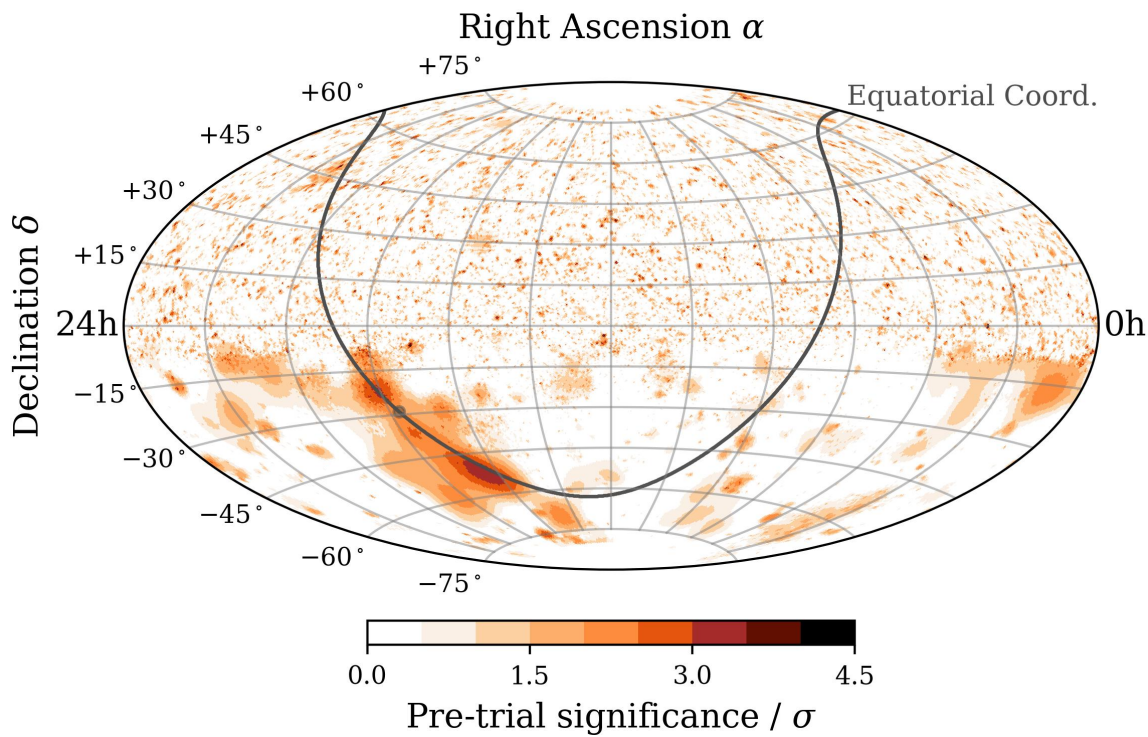}
\caption{\textbf{Combined Tracks and Cascades Skymap.} Best-fit pre-trial significance all-sky Aitoff projection map using combined cascades and tracks as a function of direction in equatorial coordinates (J2000 equinox). The solid gray line denotes the galactic plane with the dot representing the galactic center. The hottest northern spot matches the hottest northern spot seen in the tracks skymap. The hottest southern spot is not the hottest southern spot in either component skymaps. The combined skymap hottest spots are described in Table \ref{tab:all_sky_results}.
\label{fig:com_all_sky_scan}}
\end{figure}

The TS is maximized using the above-mentioned likelihood and TS on a grid of points across the entire sky (with each pixel having an approximate resolution of 0.46$^\circ$). \verb|Healpy| \citep{healpy}, a Python \citep{python} implementation of the Hierarchical Equal Area isoLatitude Pixelation (\verb|HEALPix|\footnote{\url{http://healpix.sourceforge.net}}) \citep{healpix}, was used to construct the grid by determining the pixel size with the parameter $N_\mathrm{side}=128$.
At each grid point, the local pre-trial $p$-value—that is, the probability of observing a TS value as large or larger than the one measured purely from background—is determined by comparing the observed TS to a $\chi^2$ fit of the TS distribution from multiple background trials. The point with the smallest $p$-value in each part of the sky is identified as the ``hottest spot". To calculate the post-trial probability, the $p$-value of this hottest spot is compared against the distribution of hottest spots obtained from numerous background trials for that part of the sky. If the significance of the post-trial $p$-value does not meet the $3\sigma$ evidence potential threshold, it is not high enough to claim either discovery or inconsistency with the background-only hypothesis, and thus the hottest spot is considered consistent with background emission. Each hottest spot's best fit $\hat{n}_s$, $\hat{\gamma}$ for an assumed single power-law energy spectrum, $\delta$ in degrees, and $\alpha$ in degrees is shown in Table~\ref{tab:all_sky_results}. This search excludes points that are within 10° of either celestial pole due to low quality statistics. The full pre-trial significance skymap for tracks and cascades can be seen in Fig.~\ref{fig:sep_all_sky_scan} and for combined tracks and cascades can be seen in Fig.~\ref{fig:com_all_sky_scan}.

The most significant pixel in the northern sky is found in equatorial coordinates (J2000) at a $\delta$ of $-0.30^\circ$ and $\alpha$ of 40.8$^\circ$. Accounting for trials across the whole part of the sky tested, the post-trial $p$-value at this location is 0.18. This point lies in the vicinity of (about $0.3^\circ$ away from) the galaxy NGC 1068. The best-fit $n_s$ for this point is 89 and the best-fit $\gamma$ is 3.3. The most significant pixel in the southern sky is found at a declination of $-6.88^\circ$ and $\alpha$ of 254.5$^\circ$ with a post-trial $p$-value of 0.08. The best-fit $n_s$ for this point is 73 and the best-fit $\gamma$ is 2.4. 3$^\circ$ by 3$^\circ$ scans of the region around the hottest pixel in the northern and southern sky are shown in Fig.~\ref{fig:both_hotspot_zooms} with the location of NGC 1068 in the northern sky marked with a star. Both hottest spots are consistent with the background-only hypothesis.

\begin{table}[!htb]
\centering
\small
\caption{\textbf{All-sky Search Most Significant Locations.} Summary of location, number of signal events, ($n_s$), power-law spectral index, ($\gamma$), local $p$-value, and global $p$-value of the most significant point in the northern and southern sky for tracks, cascades \citep{DNNCascades_2023}, and the combined dataset. The global $p$-value is calculated by correcting for testing locations across the source's corresponding part of the sky (either northern or southern). \label{tab:all_sky_results}}
\begin{tabular}{lccccccc}
\hline
\hline
Dataset & Hemisphere & $\alpha$[°] & $\delta$[°] & $n_s$ & $\gamma$ & $p$-value$_{\mathrm{local}}$ & $p$-value$_{\mathrm{global}}$ \\
\hline
Tracks & North & 41 & -0.3 & 69 & 3.3 & 5.37 $\times$ $10^{-6}$ & 0.22\\
{} & South & 112 & -57.4 & 27 & 1.9 & 1.55 $\times$ $10^{-6}$ & 0.05\\
\hline
Cascades & North & 338 & 17.6 & 214 & 3.6 & 3.92 $\times$ $10^{-4}$ & 0.28\\
{} & South & 248 & -50.9 & 90 & 2.9 & 1.31 $\times$ $10^{-3}$ & 0.46\\
\hline
Combined & North & 41 & -0.3 & 89 & 3.3 & 3.98 $\times$ $10^{-6}$ & 0.18\\
{} & South & 255 & -6.9 & 73 & 2.4 & 9.96 $\times$ $10^{-6}$ & 0.08\\
\hline
\end{tabular}
\end{table}

\subsubsection{Source Catalog Searches}
\begin{figure}[h!]
\includegraphics[width=\linewidth]{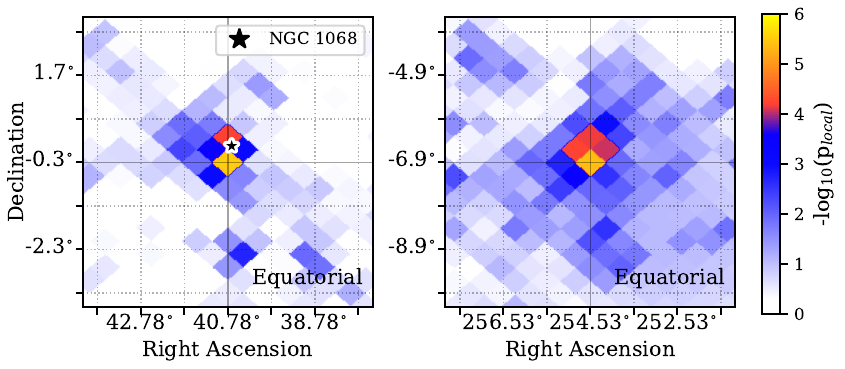}
\caption{\textbf{Northern and Southern Hottest Spots.} Local pre-trial $p$-value maps in equatorial coordinates of the area around the most significant point in the northern sky (left) and southern sky (right). The only source-list source seen over both zoomed in maps is NGC 1068. Its location is shown as the black star in the area of the northern hottest spot (left). 
\label{fig:both_hotspot_zooms}}
\end{figure}

A search on previously observed $\gamma$-ray sources is conducted using the combined tracks and cascades dataset. The catalog used in this work is comprised of 167 sources, constructed by combining the source catalogs from \cite{10yrPSTracks} and \cite{DNNCascades_2023}. The individual catalogs themselves were compiled from the fourth Fermi Large Area Telescope catalog \citep{FermiLAT_Cascades}. 103 of these sources are located in the northern sky while the rest of the 64 sources are located in the southern sky. The pre-trial $p$-values as well as each source's location in $\alpha$ and declination, best-fit $\gamma$, and best-fit $n_s$ are shown in Table~\ref{tab:source_list}. Each source's 90\% C.L. upper-limit flux assuming either an $E^{-2}$ or $E^{-3}$ energy spectrum is shown in Fig.~\ref{fig:all_sky_sens}. The most significant source in the catalog is the galaxy NGC 1068. In this search, NGC 1068 has a local pre-trial $p$-value of 1.3 $\times$ 10$^{-6}$ (4.7$\sigma$) calculated from 30 million background trials and global $p$-value of 2.1 $\times$ 10$^{-4}$ (3.5$\sigma$) obtained by trial-correcting for the 167 sources in the catalog. The best-fit $n_s$ and $\gamma$ for NGC 1068 are 92.3 and 3.1 respectively which are consistent with the all-sky northern hottest spot. The slightly elevated $n_s$ at the location of NGC 1068, relative to the most significant northern sky pixel, arises from the choice of pixel binning ($N_\mathrm{side}=128$) used in the all-sky scan, which does not center a pixel precisely at the coordinates of NGC 1068.  This is shown in Fig.~\ref{fig:both_hotspot_zooms} (left) where the hottest pixel is adjacent to another similarly hot pixel, with NGC 1068 located between them. These results show good agreement with previous evidence which has already shown NGC 1068 to be a likely neutrino source \citep{10yrPSTracks, NorthernTracks}. The significance seen in this work differs from \cite{10yrPSTracks} due to the added years of track data and from \cite{NorthernTracks} due to differences in track event reconstructions. 

Eight sources in the catalog have a pre-trial $p$-value $\le$ 0.01. These sources are NGC 1068, PKS 1424+240, PMN J1650-5044, GB6 J1542+6129, TXS 0506+056, G343.1-2.3, PMN J1603-4904, and MGRO J2019+37. Compared to the pre-trial $p$-values of corresponding sources from \cite{10yrPSTracks} and \cite{DNNCascades_2023}, NGC 1068, PKS 1424+240, PMN J1650-5044, G343.1-2.3, PMN J1603-4904, and MGRO J2019+37 in this analysis have smaller pre-trial $p$-values while those of GB6 J1542+6129 and TXS 0506+056 are larger. 

\begin{figure}[h!]
\includegraphics[width=\linewidth]{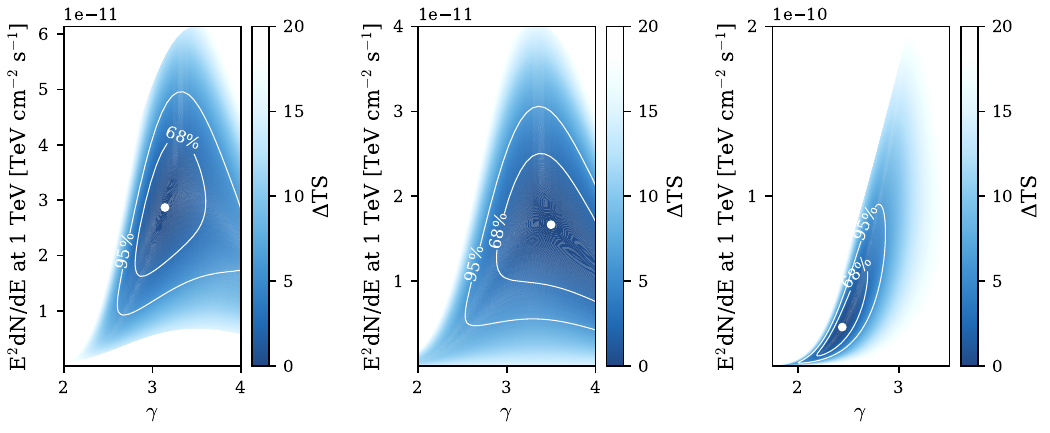}
\caption{\textbf{Likelihood Maps of NGC 1068 (left), PKS 1424+240 (middle), and the Southern Hottest Spot (right).} These maps show the likelihood contours in the per-flavor astrophysical flux as a function of spectral index ($\gamma$) with normalization at 1 TeV. 95\% and 68\% confidence level contours are shown assuming Wilks’ theorem \citep{Wilks} with two degrees of freedom. The location of the best-fit flux and best-fit spectral index are shown as white dots. 
\label{fig:likelihood}}
\end{figure}

\section{Time-Dependent Flare Search}

\subsection{Analysis Method}

IceCube follows a strict blindness procedure and the criteria for the time-dependent flare search in this work was decided prior to unblinding. The criteria for sources used in this search was the hottest northern and southern points and source list items with a time-integrated pre-trial significance of $\ge$~$3\sigma$. The source lists items which met this criteria were the NGC 1068 and PKS 1424+240. Since the hottest northern point is spatially consistent with NGC 1068, the location of the hottest northern point is substituted with the location of NGC 1068 for this flare search. Although in the top five source list items, TXS 0506+056 did not meet this criteria. IceCube does not allow unblinding of sources ad hoc after an analysis is complete.

To evaluate whether these sources exhibit time-dependent behavior when combining tracks and cascades, this analysis employs the unbinned maximum likelihood technique outlined in \cite{TXS_Flare} during the time period where both tracks and cascades overlap. The maximum allowed half-width flare duration is restricted to that of tracks selection seen in Table \ref{tab:datasamples}, however, since the cascades data sample is fully contained temporally within the tracks sample, the flare mean time is scanned only over the cascade livetime. 

The time-integrated likelihood space in $\gamma$ and $n_s$ at the coordinates of NGC 1068, PKS 1424+240, and the southern hottest spot are shown in Fig.~\ref{fig:likelihood}. The contours in Fig.~\ref{fig:likelihood} are derived assuming Wilks' theorem~\citep{Wilks} with two degrees of freedom. This assumption was validated using scrambled data as background estimations which, with the constraint that $n_s\ge0$, resulted in TS distributions that were consistent with a $\chi^2$ distribution within statistical uncertainties. This time-dependent search is done primarily to check for the possibility of a single strong flare being responsible for the time-integrated result and does not characterize the time variability of a source in detail. For this search, we assume a Gaussian distribution of events in time and use the following likelihood to search for the single most significant flare within the allowed lifetime of the source.

\begin{equation} \label{TD_likelihood}
\mathcal{L}(n_s,\gamma, T_0, \sigma_t) = \prod_{j}^{M} \prod_{i \in j}^{N}\frac{n_s^j}{N^j}\mathcal{S} \times \mathcal{T_S}(T_0, \sigma_t)+ (1 - \frac{n_s^j}{N^j})\mathcal{B}\times \mathcal{T_B}
\end{equation}

The $\mathcal{S}$ and $\mathcal{B}$ used in the time-dependent search have the same form as described in Eqs. \ref{S_gaussian} and \ref{S_vMF} for $\mathcal{S}$ and Eqn. \ref{B} for $\mathcal{B}$. However, we must now add a time component to this likelihood. The signal time PDF ($\mathcal{T_S}$) is:

\begin{equation} \label{TD_S}
\mathcal{T_S}(T_0, \sigma_t) = \frac{1}{\sqrt{2\pi\sigma_t^2}}e^{-\frac{(t_i-T_0)^2}{2\sigma_t^2}}
\end{equation}

$T_0$ is the mean time of the Gaussian flare, $\sigma_t$ is the Gaussian half-width of the flare, and $t_i$ is the event time. The background time PDF ($\mathcal{T_B}$) is approximated by $1/T_{tot}$ where $T_{tot}$ is the total observation time of the sample. 

The test statistic is calculated in almost the same way as in the time-integrated search, except now a term is added to account for the look-elsewhere effect due to the choice of $\sigma_t$ within a total observation time of $T_{tot}$:

\begin{equation} \label{TD_TS}
TS = -2 \log\left[\frac{T_{tot}}{\hat{\sigma}_t}\times \frac{\mathcal{L}(n_s=0)}{\mathcal{L}(\hat{n}_s, \hat{\gamma}, \hat{\sigma}_t, \hat{T}_0)}\right]
\end{equation}

We set the maximum allowed $\sigma_t$ to be equal to half of the dataset livetime, about 1960 days, so that the best-fit single flare cannot be longer than the total livetime. 

\subsection{Sensitivity and Discovery Potential}

\begin{figure}[h!]
\includegraphics[width=\linewidth]{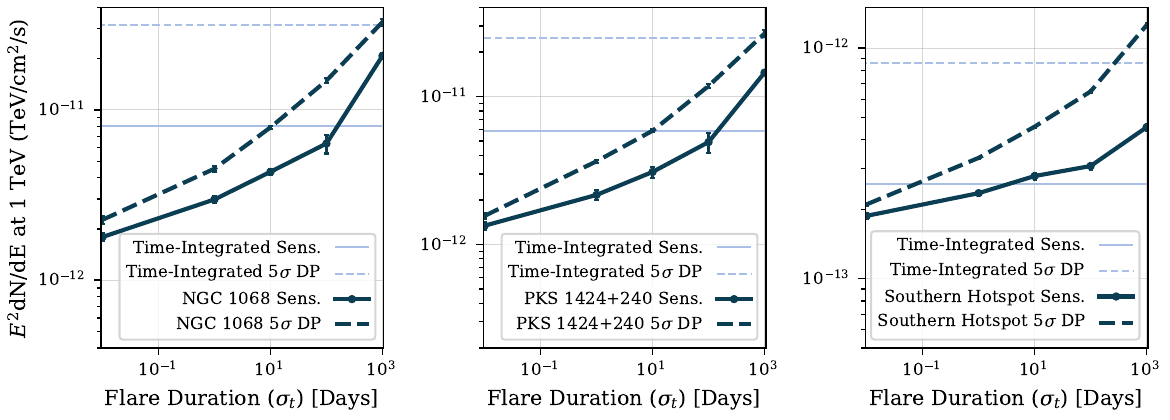}
\caption{\textbf{Time-Dependent Sensitivities and Discovery Potential.} 90\% C.L. median sensitivity to NGC 1068, PKS 1424+240, and the southern hottest spot assuming an $E^{-3}$,  $E^{-3}$, and $E^{-2}$ spectrum respectively as a function of flare duration in days for combined tracks and cascades along with the sensitivity from the time-integrated all-sky search at each source’s declination. The 5$\sigma$ discovery potential (DP) for each source is also shown along with the 5$\sigma$ DP from the time-integrated all-sky search at each source’s declination. $\frac{dN}{dE}$ is the per-flavor number of neutrinos ($N$) per neutrino energy ($E$) per area per time. 
\label{fig:fluences}}
\end{figure}

The time-dependent sensitivities and 5$\sigma$ discovery potentials at the locations of the time-integrated southern hottest spot, NGC 1068, and PKS 1424+240 are shown in Fig. \ref{fig:fluences} compared to the time-integrated sensitivity and 5$\sigma$ discovery potential at each location's declination. At shorter flare durations, the sensitivity of the flare-search is better than that of the time-integrated search since at short flare durations there is less atmospheric background contamination. At longer flare durations, the time-dependent flare-search becomes less sensitive than the time-integrated search. This occurs due to the flare-search fitting for four free parameters instead of two as in the time-integrated search. 

\subsection{Results}

\begin{figure}[h!]
\centering
\includegraphics[width=.7\linewidth]{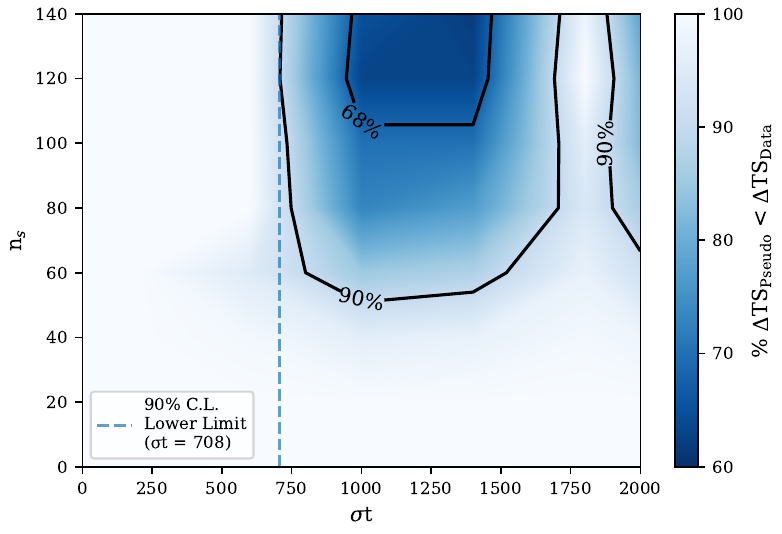}
\caption{\textbf{NGC 1068 Feldman-Cousins Contour.} This contour is constructed by comparing the change in TS from pseudo-experiments to the change in TS from data at varying points in the 4D time-dependent likelihood parameter space. The color bar represents the percentage of pseudo-experiments which exhibited smaller $\Delta$TS than data. Regions where this value are $\ge$90 are excluded. The dashed line indicates 90\% contour’s minimum $\sigma$t and thus the lower-limit on the flare time of NGC 1068. 
\label{fig:FC}}
\end{figure}

Our analysis found no evidence for flaring activity at the locations of NGC 1068, PKS 1424+240, and the southern hottest spot after trial corrections. At each location, the best-fit flare is trial-corrected by 167 from searching for flares from the catalog of 167 sources. The 90\% C.L. upper-limit fluxes ($E^2\cdot \mathrm{dN/dE}$) at 1TeV for the flares at the locations of NGC 1068, PKS 1424+240, and the southern hottest spot are $5.97\times 10^{-12}$, $5.79\times 10^{-13}$, and $5.76\times 10^{-13}$ respectively in units of $\mathrm{TeV/cm}^2\mathrm{/s}$. Although the best-fit flare at NGC 1068 exhibits a trial-corrected significance of 4.8$\sigma$, thereby rejecting the background hypothesis, its best-fit $\sigma_t$ parameter reached the maximum allowed value. To investigate further, we simulated pseudo-experiments with an injected steady NGC 1068-like signal and fitt{ed} for a best-fit flare to see how often a known steady source could cause a long-duration high-significance flare. These tests indicate that such a significance is obtained 17\% of the time. This shows the limitation of distinguishing long flares against steady emissions using this method. This limitation could be due to the higher number of four free parameters in the time-dependent Gaussian flare likelihood compared to only two free parameters in the time-integrated likelihood. {To further investigate, we implemented the Feldman-Cousins method \citep{feldman} to compare the change in test statistic, $\Delta$TS, across our 4D parameter space. We first generated signal trials by injecting simulated signals with varying combinations of $n_s$, $\gamma$, $T_0$, and $\sigma_t$. For each trial, we calculated the difference between the best-fit TS and the TS at the injected parameters to produce a distribution of $\Delta$TS$_{\mathrm{Pseudo}}$ values. From the real data, we similarly computed a single value, $\Delta$TS$_{\mathrm{Data}}$, as the difference between the best-fit TS and the TS evaluated at the same injection parameters. For each combination of parameters, $\Delta$TS$_{\mathrm{Data}}$ was then compared to the corresponding distribution of $\Delta$TS$_{\mathrm{Pseudo}}$ values, allowing us to calculate the fraction of pseudo-experiments for which $\Delta$TS$_{\mathrm{Pseudo}} < \Delta$TS$_{\mathrm{Data}}$. These results are summarized in a $n_s$ vs. $\sigma$t 2D histogram of the maximum fractions per parameter combination shown in Fig~\ref{fig:FC}, where regions at which the fraction exceeded 90\% are excluded. From this test,} we derive a lower limit of 708 days for NGC 1068’s best-fit $\sigma_t$. Therefore, we exclude any single Gaussian flare under 708 days long as the cause of the statistically significant time-integrated observation of NGC 1068. We are unable to distinguish flares longer than 708 days from a steady source. Overall, these results suggest that our current method for searching for transient astrophysical neutrino sources lacks sensitivity to very long-duration flares at known hottest spots. 

\section{Conclusion}

By combining 14 years of tracks and 10 years of cascades data, we have constructed a robust dataset that utilizes the benefits of both detection channels. This dataset provides the best all-sky time-integrated sensitivity to point-sources compared to using either tracks or cascades individually. The effective area to PeV-scale neutrinos in the southern sky is two times better than either tracks or cascades individually. The most significant pixel in the northern sky from the all-sky time-integrated point-source search lines up with NGC 1068 while the most significant pixel in the southern sky does not match any source on the source-list and is not the most significant pixel in the southern sky in either tracks or cascades individually. Although it is a newly seen hottest spot, after trial corrections, the southern hottest spot rejects the background hypothesis at a level of 1.4$\sigma$. The hottest source list item is NGC 1068 which rejects the background hypothesis at 3.5$\sigma$ after {catalog-based }post-trial corrections. Although no single-flaring activity was discovered at the locations of the southern hottest spot, NGC 1068, and PKS 1424+240, this is the first time a single-flare with duration less than four years is excluded at as being responsible for NGC 1068's emergence as a neutrino source.  

\section*{Acknowledgments}
The IceCube collaboration acknowledges the significant contributions to this manuscript from Riya Shah. The authors gratefully acknowledge the support from the following agencies and institutions:
USA {\textendash} U.S. National Science Foundation-Office of Polar Programs,
U.S. National Science Foundation-Physics Division,
U.S. National Science Foundation-EPSCoR,
U.S. National Science Foundation-Office of Advanced Cyberinfrastructure,
Wisconsin Alumni Research Foundation,
Center for High Throughput Computing (CHTC) at the University of Wisconsin{\textendash}Madison,
Open Science Grid (OSG),
Partnership to Advance Throughput Computing (PATh),
Advanced Cyberinfrastructure Coordination Ecosystem: Services {\&} Support (ACCESS),
Frontera and Ranch computing project at the Texas Advanced Computing Center,
U.S. Department of Energy-National Energy Research Scientific Computing Center,
Particle astrophysics research computing center at the University of Maryland,
Institute for Cyber-Enabled Research at Michigan State University,
Astroparticle physics computational facility at Marquette University,
NVIDIA Corporation,
and Google Cloud Platform;
Belgium {\textendash} Funds for Scientific Research (FRS-FNRS and FWO),
FWO Odysseus and Big Science programmes,
and Belgian Federal Science Policy Office (Belspo);
Germany {\textendash} Bundesministerium f{\"u}r Bildung und Forschung (BMBF),
Deutsche Forschungsgemeinschaft (DFG),
Helmholtz Alliance for Astroparticle Physics (HAP),
Initiative and Networking Fund of the Helmholtz Association,
Deutsches Elektronen Synchrotron (DESY),
and High Performance Computing cluster of the RWTH Aachen;
Sweden {\textendash} Swedish Research Council,
Swedish Polar Research Secretariat,
Swedish National Infrastructure for Computing (SNIC),
and Knut and Alice Wallenberg Foundation;
European Union {\textendash} EGI Advanced Computing for research;
Australia {\textendash} Australian Research Council;
Canada {\textendash} Natural Sciences and Engineering Research Council of Canada,
Calcul Qu{\'e}bec, Compute Ontario, Canada Foundation for Innovation, WestGrid, and Digital Research Alliance of Canada;
Denmark {\textendash} Villum Fonden, Carlsberg Foundation, and European Commission;
New Zealand {\textendash} Marsden Fund;
Japan {\textendash} Japan Society for Promotion of Science (JSPS)
and Institute for Global Prominent Research (IGPR) of Chiba University;
Korea {\textendash} National Research Foundation of Korea (NRF);
Switzerland {\textendash} Swiss National Science Foundation (SNSF).

\bibliography{paper}{}
\bibliographystyle{aasjournal}

\input{appendix}

\end{document}

%% file: appendix.tex
\clearpage
\appendix

\section{Event Selection}
Events are initially determined through IceCube's trigger system which searches for a local clustering of 8 events within a sliding time window of 5~$\mu$s~\citep{Instrumentation}. However, trigger criteria alone are insufficient for astrophysical neutrino searches, as they select events at a rate of approximately 2.7 kHz with atmospheric muons and atmospheric neutrinos outnumbering astrophysical neutrinos by roughly $10^8$:1, necessitating various event selection procedures to achieve good signal purity. 
\subsection{Tracks}
Track-like events are initially identified through the aforementioned triggers, with subsequent reconstruction using techniques such as modified least squares fits to linear trajectories and likelihood-based methods that model expected light patterns in the detector. Further selection of track events differs by hemisphere, separated by IceCube's operational horizon at a declination of $-5^\circ$ in equatorial coordinates. In the northern sky, atmospheric muons are strongly suppressed by selecting high-quality track-like events and applying a multivariate Boosted Decision Tree (BDT)~\citep{10yrPSTracks}. The BDT separates neutrino-induced tracks from atmospheric muons and cascades by selecting for high quality metrics in various track fits. The BDT retains $\approx$90\% of atmospheric neutrinos while reducing atmospheric muons to $\approx$0.1\% of their initial rate. In the southern sky, the background of atmospheric muons is orders of magnitude higher. Therefore, selection relies on strict reconstruction-quality cuts and an energy threshold of $\ge$10 TeV to suppress the softer atmospheric spectrum while retaining sensitivity to the harder astrophysical component~\citep{10yrPSTracks}. After all selection criteria are implemented, the tracks sample yields an event rate of $\sim$3 mHz, dominated by atmospheric neutrinos in the northern sky and by high-energy, well-reconstructed muons in the southern sky. 
\subsection{Cascades}
Selection for cascade events is performed using convolutional neural networks (CNNs) applied in multiple stages~\citep{DNNCascades_2023}. Early, fast CNNs reduce the atmospheric background by reconstructing cascade direction, energy, and vertex and discarding events that do not start in or near the detector volume. This fast CNN reduces the atmospheric background by $\sim$99.92\% compared to the online cascade filter, while retaining $>$50\% of neutrino-induced events above 500 GeV. Later stages employ deeper networks and boosted decision trees trained on CNN outputs, achieving $\sim$8 orders of magnitude suppression of atmospheric muons. The final cascade sample has an event rate of $\sim$0.2 mHz and contains cascade events which are fully and partially contained within the detector volume.

\clearpage
\section{A-Priori Catalog Results}
\begin{table}[!htb]
\centering
\caption{\textbf{Catalog of Sources and Results} Summary of location, number of signal events, $n_s$, power-law spectral index, $\gamma$, $-$log$_{10}$(local $p$-value) of each source in the catalog for the combined tracks and cascades ($-$log$_{10}$(p$_{\mathrm{local,\ combined}}$) and for tracks alone ($-$log$_{10}$(p$_{\mathrm{local,\ tracks}}$). A dash denotes a source with a test statistic of zero, and blank entries in the track‑only p‑value column indicate sources not included in the tracks‑only catalog. \label{tab:source_list}}
\small
\begin{tabular}{lcccccccc}
\hline
\hline
 & \textbf{Source Name} & \textbf{Class} & \textbf{$\alpha$ [°]} & \textbf{$\delta$ [°]} & \textbf{$\hat{n}_s$} & \textbf{$\hat{\gamma}$} & \textbf{$-$log$_{10}$(p$_{\mathrm{local,\ combined}}$)} & \textbf{$-$log$_{10}$(p$_{\mathrm{local,\ tracks}}$)} \\
\hline
1 & NGC 1068 & SBG & 40.67 & $-$0.01 & 92.32 & 3.14 & 5.902 & 5.685\\
2 & PKS 1424+240 & BLL & 216.76 & 23.8 & 75.79 & 3.50 & 3.534 & 3.866\\
3 & PMN J1650-5044 & BLL & 252.59 & $-$50.75 & 88.46 & 2.93 & 2.856 \\
4 & GB6 J1542+6129 & BLL & 235.75 & 61.5 & 50.79 & 3.34 & 2.709 & 2.497\\
5 & TXS 0506+056 & BLL & 77.36 & 5.7 & 9.67 & 0.87 & 2.663 & 2.793\\
6 & G343.1-2.3 & PWN & 257.0 & $-$44.3 & 77.60 & 2.92 & 2.085 \\
7 & PMN J1603-4904 & BLL & 240.97 & $-$49.06 & 84.59 & 3.10 & 2.046 \\
8 & MGRO J2019+37 & GAL & 304.85 & 36.8 & 43.35 & 2.99 & 2.000 & 1.749\\
9 & 4C +55.17 & FSRQ & 149.42 & 55.38 & 42.38 & 3.05 & 1.835 & 1.670\\
10 & M 31 & SBG & 10.82 & 41.24 & 31.83 & 4.00 & 1.786 & 1.760\\
11 & Galactic Center & BCU & 266.41 & $-$29.0 & 60.42 & 2.71 & 1.603 \\
12 & TXS 1714-336 & BLL & 259.4 & $-$33.7 & 67.69 & 2.85 & 1.563 \\
13 & PKS 1717+177 & BLL & 259.81 & 17.75 & 37.81 & 3.88 & 1.518 & 1.520 \\
14 & PKS 1830-211 & FSRQ & 278.41 & $-$21.06 & 70.55 & 2.76 & 1.484 \\
15 & PMN J1802-3940 & FSRQ & 270.67 & $-$39.67 & 76.90 & 3.12 & 1.468 \\
16 & B2 1520+31 & FSRQ & 230.55 & 31.74 & 8.12 & 1.41 & 1.447 & 0.637 \\
17 & OJ 014 & BLL & 122.86 & 1.78 & 38.29 & 4.00 & 1.438 & 1.450 \\
18 & GRS 1285.0 & UNIDB & 283.15 & 0.69 & 35.76 & 3.72 & 1.358 & 1.344 \\
19 & MGRO J1908+06 & GAL & 287.17 & 6.18 & 5.28 & 1.68 & 1.266 & 1.149 \\
20 & PKS 0048-09 & BLL & 12.68 & $-$9.49 & 126.98 & 3.59 & 1.258 & 0.622 \\
21 & KUV 00311-1938 & BLL & 8.4 & $-$19.36 & 109.29 & 3.64 & 1.257 \\
22 & AP Librae & BLL & 229.43 & $-$24.37 & 7.88 & 0.50 & 1.085 \\
23 & 4C +14.23 & FSRQ & 111.33 & 14.42 & 29.58 & 2.89 & 1.038 & 0.876\\
24 & PKS 1622-253 & FSRQ & 246.45 & $-$25.46 & 37.62 & 2.65 & 0.990 \\
25 & PKS B1130+008 & BLL & 173.2 & 0.58 & 3.93 & 1.34 & 0.955 & 0.861\\
26 & MG4 J200112+4352 & BLL & 300.3 & 43.89 & 20.84 & 2.28 & 0.933 & 0.770\\
27 & MG1 J123931+0443 & FSRQ & 189.89 & 4.73 & 5.51 & 0.80 & 0.911 & 0.885\\
28 & 3C 273 & FSRQ & 187.27 & 2.05 & 7.29 & 1.37 & 0.887 & 0.796\\
29 & 4C +28.07 & FSRQ & 39.47 & 28.8 & 3.34 & 0.50 & 0.884 & 1.038\\
30 & NGC 598 & SBG & 23.52 & 30.62 & 27.38 & 4.00 & 0.866 & 0.894\\
31 & NGC 5380 & GAL & 209.33 & 37.5 & 18.75 & 3.14 & 0.854 \\
32 & PKS 1441+25 & FSRQ & 220.99 & 25.03 & 5.82 & 2.07 & 0.818 & 0.782\\
33 & PKS 2326-502 & FSRQ & 352.33 & $-$49.93 & 5.84 & 1.77 & 0.810 \\
34 & PKS 0208-512 & FSRQ & 32.69 & $-$51.02 & 6.03 & 1.96 & 0.777 \\
35 & Gamma Cygni & GAL & 305.56 & 40.26 & 17.04 & 4.00 & 0.770 & 0.774\\
36 & CTA 102 & FSRQ & 338.15 & 11.73 & 2.09 & 0.88 & 0.757 & 0.782 \\
37 & S4 1749+70 & BLL & 267.15 & 70.1 & 6.00 & 0.50 & 0.748 & 0.706\\
38 & PSR B0656+14 & GAL & 104.95 & 14.24 & 23.34 & 4.00 & 0.742 & 0.733\\

\hline
\end{tabular}
\end{table}

\begin{table}[!htb]
\centering
\small
\begin{tabular}{lcccccccc}
\hline
\hline
 & \textbf{Source Name} & \textbf{Class} & \textbf{$\alpha$ [°]} & \textbf{$\delta$ [°]} & \textbf{$\hat{n}_s$} & \textbf{$\hat{\gamma}$} & \textbf{$-$log$_{10}$(p$_{\mathrm{local,\ combined}}$)} & \textbf{$-$log$_{10}$(p$_{\mathrm{local,\ tracks}}$)} \\
\hline
39 & TXS 1902+556 & BLL & 285.8 & 55.68 & 19.44 & 4.00 & 0.742 & 0.714\\
40 & B2 0218+357 & FSRQ & 35.28 & 35.94 & 20.10 & 4.00 & 0.719 & 0.728\\
41 & 1H 1914-194 & BLL & 289.44 & $-$19.36 & 35.35 & 2.70 & 0.714 \\
42 & BL Lac & BLL & 330.69 & 42.28 & 17.78 & 3.54 & 0.709 & 0.740\\
43 & PKS 1244-255 & FSRQ & 191.69 & $-$25.8 & 5.67 & 1.39 & 0.706 \\
44 & TXS 0518+211 & BLL & 80.44 & 21.21 & 22.91 & 3.29 & 0.687 & 0.620\\
45 & PKS 0700-661 & BLL & 105.13 & $-$66.18 & 28.52 & 2.91 & 0.669 \\
46 & PKS 2320-035 & FSRQ & 350.88 & $-$3.29 & 8.69 & 1.11 & 0.662 & 0.635\\
47 & 4C +38.41 & FSRQ & 248.82 & 38.14 & 8.10 & 2.36 & 0.654 & 0.707\\
48 & 2HWC J2031+415 & GAL & 307.93 & 41.51 & 13.95 & 2.46 & 0.653 & 0.655\\
49 & MG2 J201534+3710 & FSRQ & 303.89 & 37.18 & 20.15 & 3.12 & 0.651 & 0.555\\
50 & HESS J0835-455 & PWN & 128.29 & $-$45.19 & 4.06 & 0.50 & 0.633 \\
51 & PMN J1918-4111 & BLL & 289.56 & $-$41.19 & 54.56 & 3.74 & 0.626 \\
52 & 3C 454.3 & FSRQ & 343.5 & 16.15 & 3.50 & 2.01 & 0.605 & 0.637\\
53 & Mkn 501 & BLL & 253.47 & 39.76 & 16.12 & 3.98 & 0.603 & 0.640\\
54 & PKS 0736+01 & FSRQ & 114.82 & 1.62 & 3.66 & 2.00 & 0.601 & 0.620\\
55 & HESS J1857+026 & GAL & 284.3 & 2.67 & 20.85 & 3.06 & 0.599 & 0.553\\
56 & PKS 2155-304 & BLL & 329.71 & $-$30.23 & 61.58 & 4.00 & 0.591 \\
57 & OX 169 & FSRQ & 325.89 & 17.73 & 3.02 & 1.44 & 0.586 & 0.645\\
58 & PKS 1730-13 & FSRQ & 263.26 & $-$13.09 & 45.15 & 2.83 & 0.572 \\
59 & NGC 1275 & RDG & 49.96 & 41.51 & 15.69 & 2.94 & 0.568 & 0.608\\
60 & IC 678 & GAL & 168.56 & 6.63 & 17.89 & 3.01 & 0.568 \\
61 & PMN J2250-2806 & BLL & 342.69 & $-$28.11 & 58.55 & 4.00 & 0.565 \\
62 & PMN J0948+0022 & AGN & 147.24 & 0.37 & 4.28 & 1.23 & 0.549 & 0.611\\
63 & PMN J1329-5608 & BLL & 202.27 & $-$56.12 & 40.25 & 4.00 & 0.541 \\
64 & HESS J1852-000 & GAL & 283.0 & 0.0 & 17.08 & 4.00 & 0.539 & 0.536\\
65 & B2 1215+30 & BLL & 184.48 & 30.12 & 14.71 & 3.06 & 0.519 & 0.565\\
66 & PKS 0521-36 & AGN & 80.74 & $-$36.47 & 6.63 & 2.14 & 0.507 \\
67 & LMC & GAL & 80.0 & $-$68.75 & 31.50 & 4.00 & 0.502 & 1.304\\
68 & PG 1553+113 & BLL & 238.93 & 11.19 & 12.20 & 4.00 & 0.487 & 0.508\\
69 & PKS 1329-049 & FSRQ & 203.02 & $-$5.16 & 12.74 & 2.78 & 0.483 & 0.583\\
70 & PKS 1502+106 & FSRQ & 226.1 & 10.5 & 9.38 & 2.35 & 0.478 & 0.486\\
71 & PKS 1216-010 & BLL & 184.64 & $-$1.33 & 10.15 & 4.00 & 0.475 & 0.501 \\
72 & NGC 4945 & SBG & 196.36 & $-$49.47 & 8.84 & 2.14 & 0.475 & 2.024\\
73 & TXS 2241+406 & FSRQ & 341.06 & 40.96 & 10.97 & 4.00 & 0.473 & 0.463\\
74 & S3 0458-02 & FSRQ & 75.3 & $-$1.97 & 11.46 & 4.00 & 0.465 & 0.485\\
75 & S2 0109+22 & BLL & 18.03 & 22.75 & 12.92 & 2.89 & 0.449 & 0.463\\
76 & NGC 253 & SBG & 11.9 & $-$25.29 & 52.24 & 3.79 & 0.443 & 0.598\\
77 & Cen A & RDG & 201.38 & $-$43.02 & 10.29 & 2.27 & 0.428 \\
78 & PKS 0235+164 & BLL & 39.67 & 16.62 & 11.48 & 4.00 & 0.419 & 0.436\\
79 & MG1 J021114+1051 & BLL & 32.81 & 10.86 & 2.80 & 0.88 & 0.419 & 0.458\\
80 & B3 0609+413 & BLL & 93.22 & 41.37 & 5.57 & 2.22 & 0.401 & 0.514\\
81 & PKS 2032+107 & FSRQ & 308.85 & 10.94 & 2.58 & 0.50 & 0.391 & 0.418\\
82 & Crab Nebula & GAL & 83.63 & 22.01 & 8.55 & 2.54 & 0.382 & 0.341\\
83 & PKS 0829+046 & BLL & 127.97 & 4.49 & 11.88 & 2.86 & 0.376 & 0.371\\
84 & PKS 0118-272 & BLL & 20.12 & $-$27.02 & 44.56 & 3.94 & 0.371 \\
85 & RX J1931.1+0937 & BLL & 292.78 & 9.63 & 10.00 & 3.82 & 0.367 & 0.346\\
86 & PKS 0332-403 & BLL & 53.56 & $-$40.15 & 10.76 & 2.32 & 0.346 \\

\hline
\end{tabular}
\end{table}

\begin{table}[!htb]
\centering
\small
\begin{tabular}{lcccccccc}
\hline
\hline
 & \textbf{Source Name} & \textbf{Class} & \textbf{$\alpha$ [°]} & \textbf{$\delta$ [°]} & \textbf{$\hat{n}_s$} & \textbf{$\hat{\gamma}$} & \textbf{$-$log$_{10}$(p$_{\mathrm{local,\ combined}}$)} & \textbf{$-$log$_{10}$(p$_{\mathrm{local,\ tracks}}$)} \\
\hline
87 & OT 081 & BLL & 267.88 & 9.65 & 4.93 & 2.62 & 0.344 & 0.307\\
88 & PKS 0336-01 & FSRQ & 54.88 & $-$1.78 & 6.17 & 4.00 & 0.339 & 0.333\\
89 & PKS 0301-243 & BLL & 45.86 & $-$24.12 & 44.09 & 3.61 & 0.336 \\
90 & PKS 1936-623 & BLL & 295.35 & $-$62.18 & 26.49 & 4.00 & 0.335 \\
91 & PKS 2023-07 & FSRQ & 306.42 & $-$7.59 & 23.47 & 3.26 & 0.332 \\
92 & OJ 287 & BLL & 133.71 & 20.12 & 2.59 & 2.27 & 0.323 & 0.309 \\
93 & S4 1250+53 & BLL & 193.31 & 53.02 & 5.40 & 4.00 & 0.317 & 0.349\\
94 & W Comae & BLL & 185.38 & 28.24 & 5.24 & 4.00 & 0.315 & 0.336 \\
95 & PKS 1510-089 & FSRQ & 228.21 & $-$9.11 & 3.76 & 1.62 & 0.307 & 0.207\\
96 & HESS J1837-069 & GAL & 279.43 & $-$6.93 & 13.15 & 2.96 & 0.307 & 0.218\\
97 & PKS 1424-41 & FSRQ & 216.99 & $-$42.11 & 4.82 & 1.51 & 0.303 \\
98 & 3C 279 & FSRQ & 194.04 & $-$5.79 & 7.59 & 2.41 & 0.289 & 0.299\\
99 & PMN J0334-3725 & BLL & 53.56 & $-$37.43 & 10.89 & 2.45 & 0.274 \\
100 & 1ES 1959+650 & BLL & 300.01 & 65.15 & 5.26 & 2.74 & 0.271 & 0.612\\
101 & NGC 2146 & SBG & 94.53 & 78.33 & 2.27 & 1.15 & 0.268 \\
102 & HESS J1843-033 & GAL & 280.75 & $-$3.3 & 4.72 & 4.00 & 0.258 & 0.266\\
103 & 1RXS J130421.2-435308 & BLL & 196.09 & $-$43.9 & 7.13 & 2.34 & 0.258 \\
104 & PMN J0810-7530 & BLL & 122.75 & $-$75.5 & 16.52 & 4.00 & 0.247 \\
105 & PKS B1056-113 & BLL & 164.81 & $-$11.57 & 10.90 & 2.46 & 0.241 \\
106 & NVSS J190836-012 & UNIDB & 287.2 & $-$1.53 & 3.74 & 3.15 & 0.239 & 0.221\\
107 & PKS 2233-148 & BLL & 339.14 & $-$14.56 & 2.64 & 2.02 & 0.236 & 0.782\\
108 & B3 0133+388 & BLL & 24.14 & 39.1 & 1.58 & 3.75 & 0.230 & 0.233\\
109 & PKS 0735+17 & BLL & 114.54 & 17.71 & 2.51 & 3.49 & 0.226 & 0.215\\
110 & M 82 & SBG & 148.95 & 69.67 & 1.49 & 4.00 & 0.225 & 0.228\\
111 & PG 1246+586 & BLL & 192.08 & 58.34 & 0.49 & 2.81 & 0.202 & 0.223\\
112 & RGB J2243+203 & BLL & 340.99 & 20.36 & 1.07 & 4.00 & 0.197 &$-$\\
113 & NVSS J141826-023 & BLL & 214.61 & $-$2.56 & 2.07 & 4.00 & 0.196 & 0.217 \\
114 & PKS 0402-362 & FSRQ & 60.98 & $-$36.09 & 2.99 & 1.99 & 0.191 \\
115 & HESS J1849-000 & GAL & 282.26 & $-$0.02 & 0.85 & 3.75 & 0.187 & 0.185\\
116 & PMN J1610-6649 & BLL & 242.69 & $-$66.81 & 10.28 & 2.92 & 0.187 \\
117 & PKS 0440-00 & FSRQ & 70.66 & $-$0.29 & 0.25 & 2.16 & 0.186 & 0.222\\
118 & PKS 0727-11 & FSRQ & 112.58 & $-$11.69 & 9.70 & 2.63 & 0.186 & 0.297\\
119 & PMN J0531-4827 & BLL & 83.0 & $-$48.46 & 2.02 & 1.02 & 0.150 \\
120 & PMN J2345-1555 & FSRQ & 356.3 & $-$15.92 & 12.18 & 3.21 & 0.147 \\
121 & PKS 0823-223 & BLL & 126.5 & $-$22.51 & 1.72 & 2.10 & 0.144 \\
122 & PKS 1440-389 & BLL & 220.99 & $-$39.15 & 1.24 & 1.71 & 0.142 \\
123 & PKS 0537-441 & BLL & 84.71 & $-$44.09 & 0.86 & 1.99 & 0.135 \\
124 & PKS 0426-380 & BLL & 67.17 & $-$37.94 & 0.87 & 2.37 & 0.103 \\
125 & TXS 0628-240 & BLL & 97.74 & $-$24.11 & 1.17 & 2.96 & 0.093 \\
126 & 1ES 0806+524 & BLL & 122.46 & 52.31 & 0.0 &$-$&$-$&$-$\\
127 & PKS 0502+049 & FSRQ & 76.34 & 5.0 & 0.0 &$-$&$-$&$-$\\
128 & PKS 0447-439 & BLL & 72.36 & $-$43.84 & 0.0 &$-$&$-$\\
129 & 1ES 0647+250 & BLL & 102.7 & 25.06 & 0.0 &$-$&$-$&$-$\\
130 & PKS 0805-07 & FSRQ & 122.06 & $-$7.86 & 0.0 &$-$&$-$&$-$\\
131 & PKS 1124-186 & FSRQ & 171.76 & $-$18.96 & 0.0 &$-$&$-$\\
132 & OG +050 & FSRQ & 83.18 & 7.55 & 0.0 &$-$&$-$&$-$\\
133 & PKS 0454-234 & FSRQ & 74.26 & $-$23.41 & 0.0 &$-$&$-$\\
134 & MG2 J043337+2905 & BLL & 68.41 & 29.1 & 0.0 &$-$&$-$&$-$\\
\hline
\end{tabular}
\end{table}

\begin{table}[!htb]
\centering
\small
\begin{tabular}{lcccccccc}
\hline
\hline
 & \textbf{Source Name} & \textbf{Class} & \textbf{$\alpha$ [°]} & \textbf{$\delta$ [°]} & \textbf{$\hat{n}_s$} & \textbf{$\hat{\gamma}$\phantom{0.00}} & \textbf{$-$log$_{10}$(p$_{\mathrm{local,\ combined}}$)} & \textbf{$-$log$_{10}$(p$_{\mathrm{local,\ tracks}}$)} \\
\hline
135 & PKS 2052-47 & FSRQ & 314.07 & $-$47.24 & 0.0 & $-$\phantom{0.00} & $-$ & \\
136 & PKS 0422+00 & BLL & 66.19 & 0.6 & 0.0 & $-$\phantom{0.00} & $-$ & $-$ \\
137 & PKS 0420-01 & FSRQ & 65.83 & $-$1.33 & 0.0 & $-$\phantom{0.00} & $-$ & $-$ \\
138 & Mkn 421 & BLL & 166.12 & 38.21 & 0.0 & $-$\phantom{0.00} & $-$ & $-$ \\
139 & PKS 0215+015 & FSRQ & 34.46 & 1.74 & 0.0 & $-$\phantom{0.00} & $-$ & $-$ \\
140 & 4C +01.02 & FSRQ & 17.17 & 1.58 & 0.0 & $-$\phantom{0.00} & $-$ & $-$ \\
141 & PKS 0019+058 & BLL & 5.64 & 6.14 & 0.0 & $-$\phantom{0.00} & $-$ & $-$ \\
142 & HESS J1841-055 & GAL & 280.23 & $-$5.55 & 0.0 & $-$\phantom{0.00} & $-$ & $-$ \\
143 & Ton 599 & FSRQ & 179.88 & 29.24 & 0.0 & $-$\phantom{0.00} & $-$ & $-$ \\
144 & 3C 66A & BLL & 35.67 & 43.04 & 0.0 & $-$\phantom{0.00} & $-$ & $-$ \\
145 & 4C +21.35 & FSRQ & 186.23 & 21.38 & 0.0 & $-$\phantom{0.00} & $-$ & $-$ \\
146 & MH 2136-428 & BLL & 324.85 & $-$42.59 & 0.0 & $-$\phantom{0.00} & $-$ & \\
147 & PKS 1101-536 & BLL & 165.98 & $-$53.96 & 0.0 & $-$\phantom{0.00} & $-$ & \\
148 & B2 2114+33 & BLL & 319.06 & 33.66 & 0.0 & $-$\phantom{0.00} & $-$ & $-$ \\
149 & TXS 0141+268 & BLL & 26.15 & 27.09 & 0.0 & $-$\phantom{0.00} & $-$ & $-$ \\
150 & 1H 1013+498 & BLL & 153.77 & 49.43 & 0.0 & $-$\phantom{0.00} & $-$ & $-$ \\
151 & Arp 299 & SBG & 172.07 & 58.52 & 0.0 & $-$\phantom{0.00} & $-$ & \\
152 & 4C +01.28 & BLL & 164.62 & 1.56 & 0.0 & $-$\phantom{0.00} & $-$ & $-$ \\
153 & NGC 3424 & SBG & 162.91 & 32.89 & 0.0 & $-$\phantom{0.00} & $-$ & \\
154 & Arp 220 & SBG & 233.7 & 23.53 & 0.0 & $-$\phantom{0.00} & $-$ & \\
155 & PKS 1502+036 & AGN & 226.26 & 3.44 & 0.0 & $-$\phantom{0.00} & $-$ & $-$ \\
156 & 1H 1720+117 & BLL & 261.27 & 11.87 & 0.0 & $-$\phantom{0.00} & $-$ & $-$ \\
157 & SMC & SBG & 14.5 & $-$72.75 & 0.0 & $-$\phantom{0.00} & $-$ & 0.169 \\
158 & B3 1343+451 & FSRQ & 206.4 & 44.88 & 0.0 & $-$\phantom{0.00} & $-$ & $-$ \\
159 & PKS 2005-489 & BLL & 302.36 & $-$48.82 & 0.0 & $-$\phantom{0.00} & $-$ & \\
160 & PKS 2247-131 & BCU & 342.5 & $-$12.85 & 0.0 & $-$\phantom{0.00} & $-$ & \\
161 & M 87 & AGN & 187.71 & 12.39 & 0.0 & $-$\phantom{0.00} & $-$ & $-$ \\
162 & ON 246 & BLL & 187.56 & 25.3 & 0.0 & $-$\phantom{0.00} & $-$ & $-$ \\
163 & PKS 2142-75 & FSRQ & 326.83 & $-$75.6 & 0.0 & $-$\phantom{0.00} & $-$ & \\
164 & PG 1218+304 & BLL & 185.34 & 30.17 & 0.0 & $-$\phantom{0.00} & $-$ & $-$ \\
165 & S4 0814+42 & BLL & 124.56 & 42.38 & 0.0 & $-$\phantom{0.00} & $-$ & $-$ \\
166 & 1RXS J194246.3+1 & BLL & 295.7 & 10.56 & 0.0 & $-$\phantom{0.00} & $-$ & $-$ \\
167 & S5 0716+71 & BLL & 110.49 & 71.34 & 0.0 & $-$\phantom{0.00} & $-$ & $-$ \\
\hline
\end{tabular}
\end{table}

\clearpage
\section{A-Priori Catalog Construction}

\begin{figure}[htb!]
\centering
\includegraphics[width=\linewidth]{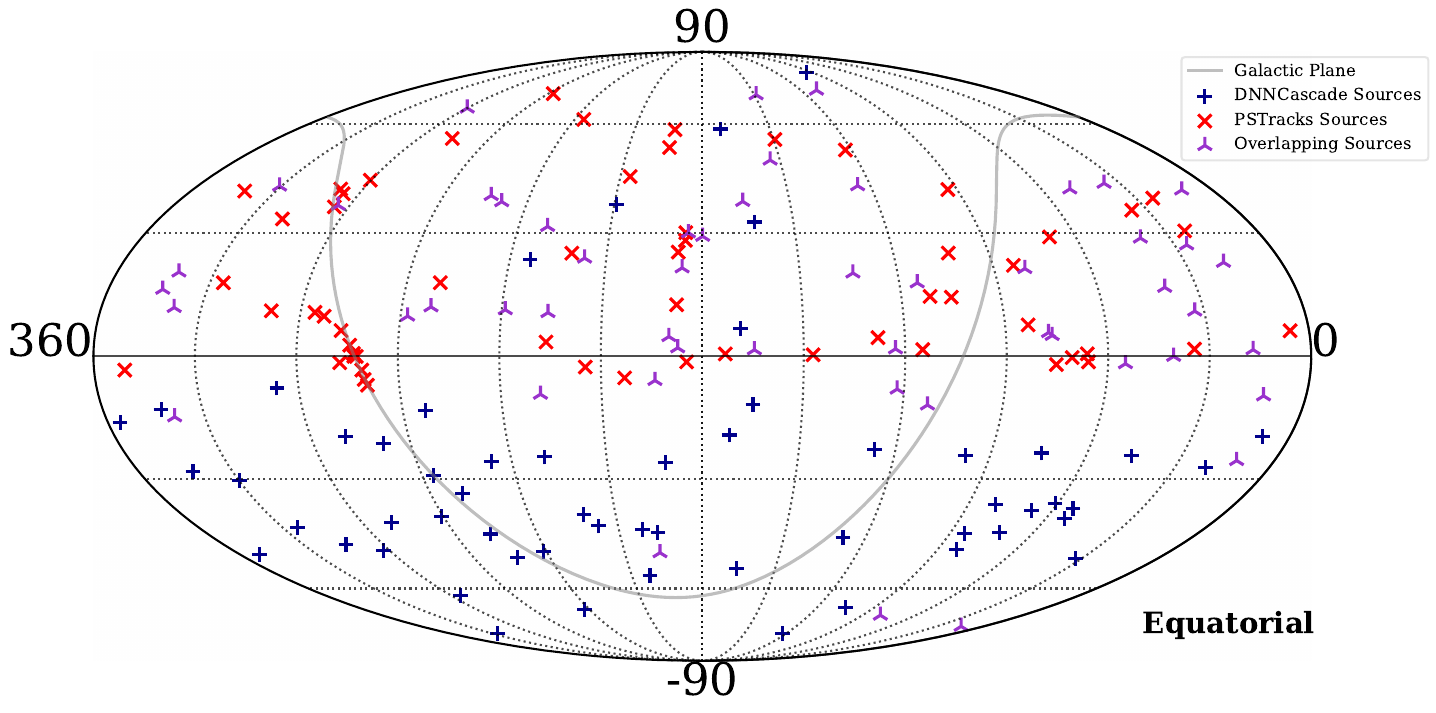}
\caption{\textbf{Source List Locations.} The locations, in Equatorial Coordinates, of the sources in the a-priori defined catalog with point-style indicating whether the source is from the previously unblinded cascades all-sky neutrino point-source \citep{DNNCascades_2023} or from the previously unblinded tracks all-sky neutrino point-source search \citep{10yrPSTracks}. 
\label{fig:source_locs}}
\end{figure}

Fig.~\ref{fig:source_locs} shows the locations of all 167 sources in the a-priori source catalog in equatorial coordinates and whether each source came from \cite{10yrPSTracks} (PSTracks Sources), \cite{DNNCascades_2023} (DNNCascade Sources), or from both (Overlapping Sources). 

\clearpage
\section{Sensitive Energy Range}
\begin{figure}[!htb]
\centering
\includegraphics[width=.9\linewidth]{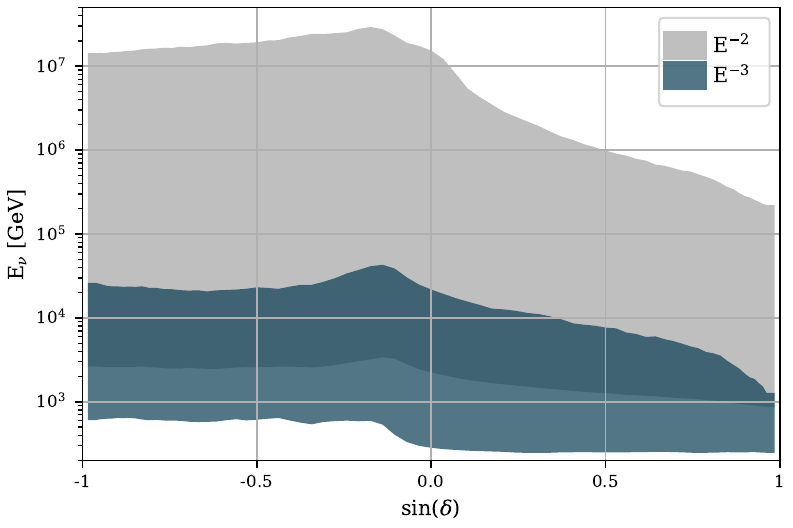}
\caption{\textbf{Sensitive Energy Range.} The central 90\% sensitive energy range of the combined tracks and cascades dataset assuming an $E^{-2}$ and $E^{-3}$ spectrum as a function of the sine of declination. The energy range is shown for declinations between -80$^{\circ}$ and +80$^{\circ}$. 
\label{fig:sens_energy}}
\end{figure}

Fig.~\ref{fig:sens_energy} shows the central 90\% sensitive energy range of the combined tracks and cascades dataset assuming an $E^{-2}$ and an $E^{-3}$ energy spectrum. In the northern sky, for an $E^{-2}$ spectrum the sensitivity spans roughly from 1 TeV up to 10 PeV near the horizon, decreasing to about 100 TeV at the northernmost declinations. For an $E^{-3}$ spectrum, the range narrows from approximately 200 GeV to 10 TeV near the horizon, tapering to around 1 TeV at the most northern declinations. In the southern sky, the ranges are broader: approximately 3 TeV to 11 PeV for an $E^{-2}$ spectrum, and about 500 GeV to 11 TeV for an $E^{-3}$ spectrum.

\clearpage
\section{Sensitivity Comparisons by Dataset}
\begin{figure}[htb!]
\centering
\includegraphics[width=.6\textwidth]{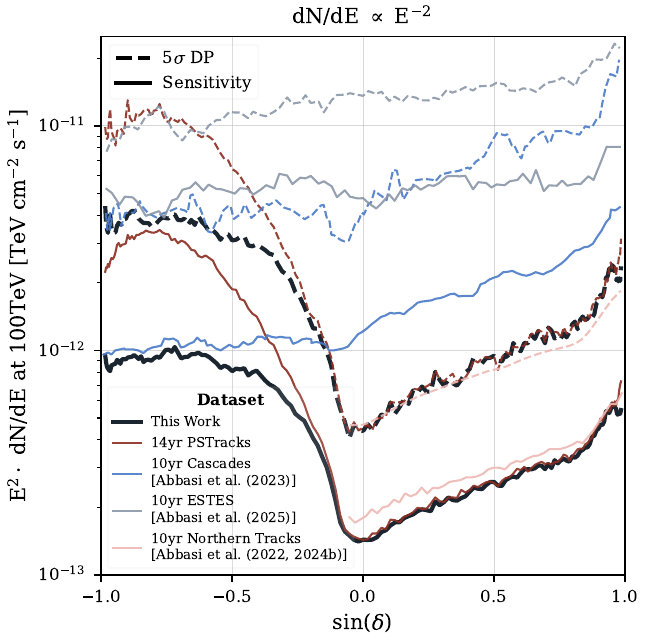}
\caption{\textbf{Sensitivity and 5$\sigma$ Discovery Potential Comparison.} 90\% C.L. median sensitivity to sources emitting an $E^{-2}$ spectrum as a function of source declination for cascades \citep{DNNCascades_2023} and PSTracks individually, the 10 year ESTES selection \citep{ESTES}, the 10 year Northern Tracks selection \citep{NorthernTracks, Seyfert2024}, and for the combined tracks and cascades are shown as solid lines. The 5$\sigma$ DP for all selections are shown as dashed lines. $\frac{dN}{dE}$ is the per-flavor number of neutrinos ($N$) per neutrino energy ($E$) per area per time.
\label{fig:all_data_sens}}
\end{figure}

Fig. \ref{fig:all_data_sens} shows the combined tracks and cascades all-sky 90\% C.L. median sensitivity and 5$\sigma$ discovery potential compared to those of various other IceCube data selections. These include the 14 year PSTracks (this work) and the 10 year cascades \citep{DNNCascades_2023}, which are both components of the combined tracks and cascades dataset. Other selections shown are the Northern Tracks selection \citep{NorthernTracks} and the ESTES selection \citep{ESTES}. The combined tracks and cascades dataset has the best all-sky sensitivity and 5$\sigma$ discovery potential compared to the individual selections shown in Fig. \ref{fig:all_data_sens}. 

